\documentclass[twocolumn,
showpacs,preprintnumbers,superscriptaddress,amsmath,amssymb,floatfix,aps,pra]{revtex4}

\usepackage[dvips]{graphicx}% Include figure files
\usepackage{dcolumn}% Align table columns on decimal point
\usepackage{bm}% bold math
\usepackage{ulem}

\begin{document}

%\preprint{APS/123-QED}

\title{Casimir-Lifshitz force out of thermal equilibrium}

\author{Mauro Antezza}
\email[Electronic address: ]{antezza@science.unitn.it}
\affiliation{Dipartimento di Fisica, Universit\`a di Trento
and CNR-INFM R\&D Center on Bose-Einstein Condensation, Via Sommarive 14, I-38050 Povo, Trento, Italy}
\author{Lev P. Pitaevskii}
\affiliation{Dipartimento di Fisica, Universit\`a di Trento and
CNR-INFM R\&D Center on Bose-Einstein Condensation, Via Sommarive
14, I-38050 Povo, Trento, Italy} \affiliation{Kapitza Institute for
Physical Problems, ul. Kosygina 2, 119334 Moscow, Russia}
\author{Sandro
Stringari}
\affiliation{Dipartimento di Fisica, Universit\`a di Trento
and CNR-INFM R\&D Center on Bose-Einstein Condensation, Via Sommarive 14, I-38050 Povo, Trento, Italy}
\author{Vitaly B. Svetovoy}
\affiliation{MESA+ Research Institute, University of Twente, PO 217, 7500 AE Enschede, The Netherlands}

\date{\today}

\begin{abstract}
We study the Casimir-Lifshitz interaction out of thermal
equilibrium, when the interacting objects are at different temperatures. The analysis is focused on the
surface-surface, surface-rarefied body, and surface-atom configurations. A systematic investigation
of the contributions to the force  coming from the propagating and
evanescent components of the electromagnetic radiation is performed.
The large distance behaviors of such interactions is discussed, and
both  analytical and numerical results are compared with
the equilibrium ones. A detailed analysis of the crossing between
the surface-surface and the surface-rarefied body, and finally the
surface-atom force is shown, and a complete derivation and
discussion of the recently predicted non-additivity effects and new
asymptotic behaviors is presented.
\end{abstract}

\maketitle

%%%%%%%%%%%%%%%%%%%%%%%%%%%%%%%%%%%%%%%%%%%%%%%%%%%%%%%%%%%%%%%%%%%%%%%%%%%%%%%%%%%%%%%%%%%%%%%%%%%%%%%%%%%%%
\section{\label{Intro}Introduction}
%%%%%%%%%%%%%%%%%%%%%%%%%%%%%%%%%%%%%%%%%%%%%%%%%%%%%%%%%%%%%%%%%%%%%%%%%%%%%%%%%%%%%%%%%%%%%
The Casimir-Lifshitz force is a {\it dispersion} interaction of
electromagnetic origin acting between neutral dispersive bodies
without permanent polarizations. The original Casimir intuition
about the presence of such a force between two parallel ideal
mirrors  \cite{Casimir} (or between an atom and a mirror, i.e. the
so called Casimir-Polder force \cite{CP}) was readily extended to
real materials by Lifshitz 
\cite{Lif56,lifshitzDAN,lifshitz}. He used the theory of
electromagnetic fluctuations  developed by Rytov \cite{Rytov} to
formulate the most general theory of the dispersion interaction in
the framework of the statistical physics and 
macroscopic electrodynamics (see also \cite{LP}). The Lifshitz theory is still the most
advanced one; today it is extensively
accepted providing a
common tool to deal with dispersive forces in different fields of science (physics, biology,
chemistry) and technology.

It is useful to stress here that the geometry of the system is
relevant for the explicit calculation of the force, but does not
affect the nature of the interaction that preserves all its
peculiar characteristics and relevant length-scales. For this reason
we refer to the Casimir-Lifshitz force for all geometrical configurations. In particular, in this paper we are interested in the force between flat and parallel
surfaces of two macroscopic bodies, and between a surface and
an individual atom.

The Lifshitz theory is formulated for systems at thermal
equilibrium. In this theory the pure quantum
effect at $T=0$ is clearly separated from
the finite temperature effect. The
former gives a dominant contribution at small separation
($<1\;\mu$m at room temperature) between the bodies and was
readily confirmed experimentally with good accuracy [see
\cite{hinds1993} (surface-atom),
\cite{lamoreaux1997,Har00,Cha01,Dec03} (surface-sphere), \cite{ruoso2002} (surface-surface)]. 

The thermal component prevails at larger distances and was measured only recently at JILA in
experiments with cold atoms \cite{Cornell06}. These experiments are based on the measurement of the shift
of the collective oscillations of a Bose-Einstein
condensate (BEC) of trapped atoms close to a surface
\cite{articolo1,eric05}. The JILA group  measured the
 Casimir-Lifshitz force at very large distances
($\sim 10 \mu$m) and for the first time showed the thermal effects of the Casimir-Lifhitz interaction (and
indeed of any dispersion interaction), in agreement with the
theoretical predictions \cite{articolo2}. This measurement was done
out of thermal equilibrium \cite{expmacrther}, where thermal effects are stronger. 

There was an interest in configurations out of thermal
equilibrium since the work by Rosenkrans {\it et al.}
\cite{Linder68} (atom-atom). Surface-atom interaction was analyzed
by Henkel {\it et al.} \cite{Henkel1} and by Antezza {\it et al.}
\cite{articolo2,AntezzaJPA,PitaevskiiJPA,ShortArticle,AntezzaPhDthesis}.
Surface-surface force was investigated by Dorofeyev {\it et al.}
\cite{Dorofeyev1,Dorofeyev2} and Antezza {\it et al.}
\cite{ShortArticle,AntezzaPhDthesis}. For a review of non-equilibrium effects see
also \cite{Greffet05}.

Further non-equilibrium effects were explored by Polder and Van
Hove \cite{Polder}, who calculated the heat-flux between two
parallel plates, and Bimonte \cite{Bimonte}, who expressed
fluctuations of fields for the metal-metal configuration in terms of
surface impedance.

The principal interest in the study of systems out of thermal
equilibrium is connected to the possibility of tuning the
interaction in both strength and sign \cite{articolo2,ShortArticle}. Such systems
give also the way to explore  the  role of thermal
fluctuations, usually masked at thermal equilibrium by the
$T=0$ component which dominates the interaction up to very large distances, where the actual total force
results to be very small.

A crucial role in explaining the peculiarity of the non-equilibrium
surface-atom force is played by cancellation
effects between the fluctuations of the different components of the
radiations, as the incident to and emitted by the surface
\cite{articolo2}.

In this paper we present a detailed study of the Casimir-Lifshitz
force out of thermal equilibrium, with particular attention devoted
to the surface-surface and surface-atom interactions. We 
perform a systematic investigation of the contributions to the force
coming from the propagating and evanescent components of the
electromagnetic radiation. The large distance behaviors of these interactions 
are extensively discussed, both analytically and numerically, and
comparisons with the equilibrium results are done. We perform a detailed
analysis of the relation between the surface-surface interaction
when one body is rarefied (surface-rarefied body force) and the
surface-atom force. We also present a complete derivation and
discussion of the recently predicted non-additivity effects and new
asymptotic behaviors noted in \cite{ShortArticle}.

We are interested in the force occurring between two planar bodies,
which are kept at different temperatures and separated by a distance
$l$. We consider that the bodies are thick enough, in order to exclude possible effects of the presence of the vacuum gap on the radiation outside the two bodies.
We also assume that each body is in \textit{local} thermal equilibrium, the whole system being in a \textit{stationary state}. In our configuration the left-side
body, $1$, has a complex dielectric function
$\varepsilon_1(\omega)=\varepsilon_1'(\omega)+i\varepsilon_1''(\omega)$,
occupies the volume $V_1$ and is held at temperature
$T_1$. The right-side body, $2$, has a complex dielectric
function
$\varepsilon_2(\omega)=\varepsilon_2'(\omega)+i\varepsilon_2''(\omega)$,
occupies the volume $V_2$ and is held at temperature
$T_2$.  First we assume that each body fills
an infinite half-space, in particular $V_1$ and $V_2$ coincide with the left and right half-spaces, respectively. Later we consider a more general situation of two 
 parallel thick slabs with the external regions shined by
the thermal radiations at arbitrary temperatures. In this case additional 
distance-independent contributions to the pressure are present. Finally we will consider the case in which one of the two bodies is rarefied. In this case the interplay between the finite thickness of the body and the non equilibrium configuration leads to different interesting behaviors of the pressure.

\begin{figure}[ptb]
\begin{center}
\includegraphics[width=0.45\textwidth]{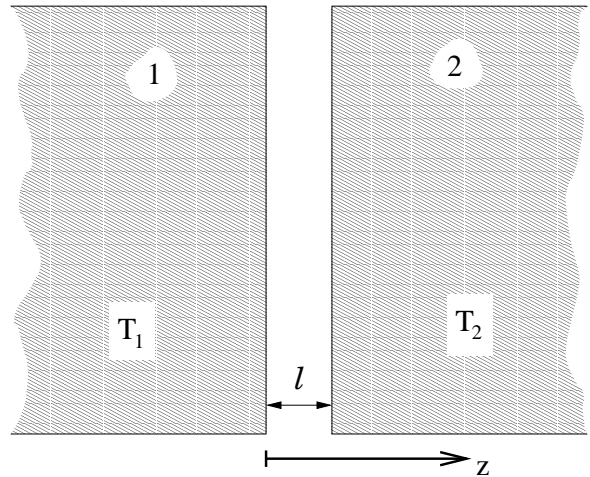}
\caption{Schematic figure of the surface-surface system out of thermal equilibrium. Here the two bodies occupy infinite half-spaces.}
\label{FiguraSSneq}
\end{center}
\end{figure}

The general problem can be set in the following way, for two bodies occupying the two half-spaces.
Let us choose the origin of the coordinate system at the boundary of
the half-space $1$ and let us set the $z$-axis in the direction of the
half-space $2$ [see Fig.\ref{FiguraSSneq}]. 
%===================================================================================================
%===================================================================================================
The electromagnetic pressure between the two
bodies along $z$ can be calculated as \cite{LLPCM,Forcebetween}
\begin{equation}
P^{\textrm{neq}}(T_1,T_2,l)=\langle T_{zz}({\bf r},t) \rangle,
\label{force}
\end{equation}
that should be regularized by subtracting the same expression at separation $l\rightarrow\infty$.
%to which  should be subtracted the same expression for $l\rightarrow\infty$. Such a subtraction has the effect to %eliminate the usual ultraviolet divergence (in the $T=0$ part) and constant terms (in the thermal part at equilibrium). 
In Eq.(\ref{force}), ${\bf r}$ is a generic point between the two bodies, and
\begin{equation}
T_{zz}({\bf r},t)=
-\frac{\Lambda_{\alpha\beta}}{8\pi}\left[E_{\alpha}({\bf
r},t)E_{\beta} ({\bf r},t)+B_{\alpha}({\bf r},t)B_{\beta}({\bf
r},t)\right],
 \label{;'vjlk}
\end{equation}
is the $zz$ component of the Maxwell stress tensor in the vacuum gap.
%%
%\begin{eqnarray}
%T_{ij}({\bf r},t)&=&
%\frac{1}{4\pi}\left\{E_{i}({\bf r},t)E_{j}({\bf r},t)+
%B_{i}({\bf r},t)B_{j}({\bf r},t)-\right.\notag\\
%&&\left.\left[{\bf E}({\bf r},t)^2+{\bf B}
%({\bf r},t)^2\right]\frac{\delta_{ij}}{2}\right\},
%\label{ojkvow}
%\end{eqnarray}
%%
%and
Here  $\Lambda_{\alpha\beta}$ is a diagonal matrix with
$\Lambda_{11}=\Lambda_{22}=1$ and $\Lambda_{33}=-1$.

To calculate the pressure (\ref{force}) one must average
over the state of the electromagnetic field the squares of the
spatial components of the electric and magnetic field
${\bf E}({\bf r},t)$ and ${\bf B}({\bf r},t)$, which appear in
Eq.~(\ref{;'vjlk}).

Before starting with the analysis of the problem we mention the structure of this work in the following outline. In  Sec. \ref{Form} we develop the
formalism, introduce the role and the description of the
fluctuations of the electromagnetic field and specify the approach
we adopt to deal with the surface optics. In Sec.
\ref{sec:LifTewwqq} we recall the main results of the surface-surface Casimir-Lifshitz
interaction at thermal equilibrium, and in particular specify the
distinction between the $T=0$ (purely quantum) and the thermal 
contribution to the force, generated by the radiation pressure of the
thermal radiation. In Sec. \ref{sec:LifTerNONEq} we present a detailed
derivation of the surface-surface pressure out of thermal
equilibrium $P^{\textrm{neq}}(T_1,T_2,l)$. In Sec.  \ref{CompPrev} we show an alternative and
useful expression for $P^{\textrm{neq}}(T_1,T_2,l)$, together with
numerical results relative to particular couples of dielectric
materials (i.e. fused silica-silicon and sapphire-fused silica).
In Sec. \ref{sec:freeedd} we deal with the distance-independent
terms in the pressure due to the finite thickness of the two bodies,
and the eventual effect of external radiation at different
temperature impinging the external surfaces. In Sec. \ref{sec:LongD}
we derive the large distance behavior of the surface-surface
pressure out of thermal equilibrium, and discuss the role of the propagating
waves (PW) and evanescent waves (EW) contributions. We
also make a comparison with the corresponding terms of the pressure
at thermal equilibrium. In Sec. \ref{sec:LongDnonad} we consider the
interaction between a surface and a rarefied body and derive the
large distance behaviors  of the PW and EW components. In the same
section we stress the presence of non additivity in the interaction
out of equilibrium (in contrast
with the equilibrium case) and show the analysis of the  crossing
between different asymptotic behaviors. In Sec. \ref{AtSurfNeq} we
show the transition from the
surface-rarefied body to surface-atom interactions out of thermal
equilibrium, and demonstrate the essential role of finite thickness of the rarefied body. Finally in Sec.
\ref{conclusions} we provide our conclusions.

In appendix   \ref{sec:GTomas} we give some
details on the expression of the Green functions we used in
our calculation, and  in appendix
\ref{sec:GLevs} we discuss in detail the force acting between a surface and a 
rarefied body of finite thickness.
%
%%%%%%%%%%%%%%%%%%%%%%%%%%%%%%%%%%%%%%%%%%%%%%%%%%%%%%%%%%%%%%%%%%%%%%%%%%%%%%%%%%%%%%%%%%%%%%
\section{\label{Form}The Formalism}
%%%%%%%%%%%%%%%%%%%%%%%%%%%%%%%%%%%%%%%%%%%%%%%%%%%%%%%%%%%%%%%%%%%%%%%%%%%%%%%%%%%%%%%%%%%%
Our approach is based on the theory of the fluctuating
electromagnetic (EM) field developed by Rytov \cite{Rytov}. In this approach it is assumed
that the field is driven by randomly fluctuating current density or,
alternatively, by randomly fluctuating polarization field. In this respect
the Maxwell equations become of Langevin-type. For a
monochromatic field  in a non-homogeneous,
linear, and nonmagnetic medium with the dielectric function
$\varepsilon(\omega,\textbf{r})$ the Maxwell equations become:
\begin{eqnarray}
\nabla\wedge{\bf E}[\omega;{\bf r}]-ik\;{\bf B}[\omega;{\bf r}]&=&0,
\label{Max1}\\
\nabla\wedge{\bf B}[\omega;{\bf r}]+ ik\;\varepsilon(\omega;{\bf
r}){\bf E}[\omega;{\bf r}]&=& -4\pi\;ik\;{\bf P}[\omega;{\bf r}],
\label{Max2}
\end{eqnarray}
where $k=\omega/c$ is the vacuum wavenumber, and $\wedge$ is the vector product symbol. The source of the
electromagnetic fluctuations is described by the electric
polarization ${\bf P}[\omega;{\bf r}]$, related to the electric
current density as ${\bf J}[\omega;{\bf r}]=-i\omega{\bf
P}[\omega;{\bf r}]$. We use the following notations for
the frequency Fourier transforms $A[\omega;{\bf r}]$ of the quantity
$A(t,{\bf r})$:
\begin{equation}
A(t,{\bf r})=\int_{-\infty}^{+\infty}\frac{\textrm{d}\omega}{2\pi}
e^{-i\omega t} A[\omega;{\bf r}]. \label{efjvih}
\end{equation}

To find the solution of the Maxwell equations we use the Green
functions formalism. A Green function is a solution of
the wave equation for a point source in presence of surrounding matter. When this solution is known
one can construct the solution due to a general source using the
principle of linear superposition. This method takes into account the effects of non-additivity, which originates from
the fact that the interaction between two fluctuating dipoles is influenced by the presence of a
third dipole. Employing this formalism we can
express the electric field at the observation point ${\bf r}$ as the
convolution
\begin{equation}
{\bf E}\left[\omega;{\bf r}\right]= \int \overline{{\bf
G}}\left[\omega;{\bf r},{\bf r}'\right]\; \cdot {\bf
P}\left[\omega;{\bf r}'\right]\;\textrm{d}{\bf r}'.
 \label{lafh}
\end{equation}
Here ${\bf P}\left[\omega;{\bf r}'\right]$ is the random polarization 
at the source point ${\bf r}'$, and $\overline{{\bf G}}\left[\omega;{\bf r},{\bf r}'\right]$ is the dyadic Green function of the system. Then it is clear that the Green function plays the role of the response function in a linear-response theory. The Green function is  the solution of 
the following equation \cite{LLP}
\begin{equation}
\left\{\nabla\wedge\nabla\wedge
-k^2\varepsilon(\omega,{\bf r})\right\}\overline{{\bf G}}[\omega;{\bf
r},{\bf r}'] =4\pi k^2\overline{{\bf I}}\delta({\bf r}-{\bf r}'),
\label{elweqcfe}
\end{equation}
where $\overline{{\bf I}}$ is the identity dyad. This equation,
resulting from the Maxwell equations (\ref{Max1}) and (\ref{Max2})
and convolution (\ref{lafh}),  has to be solved with proper boundary
conditions characterizing the fields components at the interfaces,
as well as the condition required by a retarded Green's
function \cite{retarded}, i.e. $\overline{{\bf G}}[\omega;{\bf
r},{\bf r}']\rightarrow0$ as $|{\bf r}-{\bf r}'|\rightarrow\infty$.

Finally it is useful to recall the relations $G_{\alpha\beta}[\omega;{\bf
r},{\bf r}']=G_{\beta\alpha}[\omega;{\bf r}',{\bf r}]$  and
$G_{\alpha\beta}^*[\omega;{\bf r},{\bf
r}']=G_{\alpha\beta}[-\omega;{\bf r},{\bf r}']$, that are the consequence of the microscopic reversibility in the
linear-response theory and the reality of the time dependent fields,
respectively.
%
%%%%%%%%%%%%%%%%%%%%%%%%%%%%%%%%%%%%%%%%%%%%%%%%%%%%%%%%%%%%%%%%%%%%%%%%%%%%%%%%%%%%%
\subsection{\label{FPollSec}Field correlation functions}
%%%%%%%%%%%%%%%%%%%%%%%%%%%%%%%%%%%%%%%%%%%%%%%%%%%%%%%%%%%%%%%%%%%%%%%%%%%%%%%%%%%%%
From Eq.(\ref{force}) it is evident that we are interested in the
time correlations between different components of the electric
(magnetic) field at equal times. In the quantum theory such correlations are
described by the averages of symmetrized products of the field components:
\begin{eqnarray}
\lefteqn{\!\!\!\!\!\!\left\langle E_{\alpha }(\mathbf{r},t)E_{\beta }(\mathbf{r}'
,t)\right\rangle_{\textrm{sym}}\equiv}\notag\\
&&\frac{1}{2}\left\langle E_{\alpha }(\mathbf{r}%
,t)E_{\beta }(\mathbf{r}' ,t)+E_{\beta }(\mathbf{r}' ,t)E_{\alpha
}(\mathbf{r},t)\right\rangle.
\label{oosod}
\end{eqnarray}
Notice that, although in this paper we are using symmetrized correlations, other
possible forms of the correlation functions could be more
appropriate in other situations \cite{Agarwal}. The correlations
(\ref{oosod}) in terms of their Fourier transforms can be presented
as
\begin{eqnarray}
\lefteqn{\!\!\!\!\!\!\left\langle E_\alpha(\textbf{r},t)E_\beta(\textbf{r}',t)\right\rangle_{\textrm{sym}}=}\notag\\
&&\!\!\!\!\!\!\iint\frac{\textrm{d}\omega}{2\pi}\frac{\textrm{d}\omega'}{2\pi}e^{-i(\omega-\omega')t}
\left\langle E_\alpha[\omega;\textbf{r}]\;E^{\dag}_\beta[\omega';\textbf{r}']\right\rangle_{\textrm{sym}}.
    \label{corr}
\end{eqnarray}
Using Eq.(\ref{lafh}) these correlations can be expressed via the
correlations of the polarization field $\bf P$, which obeys the
fluctuation-dissipation theorem \cite{LLPCM}
\begin{eqnarray}
\lefteqn{\!\!\!\!\!\!\left\langle
P_{\alpha}[\omega;{\textbf{r}}]P_{\beta}^{\dag}[\omega';{\textbf{r}}']\right\rangle_{\textrm{sym}}=}\notag\\
&&\!\!\!\!\!\!\frac{\hbar\varepsilon''(\omega,\textbf{r})}{2}\coth\left(\frac{\hbar\omega}{2k_B
T}\right)\delta(\omega-\omega')\delta({\bf r}-{\bf r}')\delta_{\alpha\beta},
\label{corrP}
\end{eqnarray}
expressed via the Fourier transformed
${\bf P}\left[\omega;\textbf{r}\right]$. Due to the presence of the
$\delta \left( {\bf r}-{\bf r}'\right)$ factor these fluctuations
are local.  Fluctuations of the sources in different points of the
material are non-coherent. This permits to assume that in the
non-equilibrium situation, when temperature $ T $  is different in
different points, the sources correlations are given by the same
equations. We must emphasize that this assumption, even being quite
reasonable, is still a hypothesis, which is worth both of further
theoretical investigation and experimental verification. The problem was discussed previously (see particularly \cite{Eckhardt}), but in our opinion the conditions of applicability of the theory has not been still established. The same
assumption was used by Polder and Van Hove \cite{Polder} to calculate the
radiative heat transfer between two bodies with different
temperatures.

The assumption (\ref{corrP}) (local source hypothesis)
represents the  starting point of our analysis allowing for an
explicit calculation of the electromagnetic field also if the system
is not in global thermal equilibrium.

It is now evident that EM field in the vacuum gap is given by the
sum of the fields produced by the fluctuating polarizations in the
materials filling respectively the half-space $1$,  with the
dielectric function $\varepsilon_1(\omega)$ and temperature $T_1$,
and the half-space $2$ with the dielectric function
$\varepsilon_2(\omega)$ and temperature  $T_2$. Then the Fourier transform of the
electric field correlations can be presented as
\begin{widetext}
\begin{gather}
    \left\langle E_\alpha[\omega;\textbf{r}]E_\beta^{\dag}[\omega';\textbf{r}']\right\rangle_{\textrm{sym}}=
    \left[\frac{\hbar\varepsilon''_1(\omega)}{2}\coth\left(\frac{\hbar\omega}{2k_B
T_1}\right)
    S_{\alpha\beta}^{(1)}\left[\omega;{\bf r},{\bf r}'\right]+\frac{\hbar\varepsilon''_2(\omega)}{2}\coth\left(\frac{\hbar\omega}{2k_B
T_2}\right)
    S_{\alpha\beta}^{(2)}\left[\omega;{\bf r},{\bf r}'\right]\right]\delta\left(\omega-\omega'\right),
    \label{corr2b}
\end{gather}
\end{widetext}
where $S^{(i)}_{\alpha\beta}\ (i=1,2)$ is defined as convolution of
two Green functions
\begin{equation}
    S^{(i)}_{\alpha\beta}\left[\omega;{\bf r},{\bf
    r}'\right]=\int_{V_i}d{\bf r}''G_{\alpha\gamma}[\omega;{\bf r},{\bf r}'']
    G_{\gamma\beta}^*[\omega;{\bf r}'',{\bf r}'].
    \label{Sdef}
\end{equation}
Here $V_1$ and $V_2$ are the volumes occupied by the left and right
body, respectively, and the two terms in Eq.(\ref{corr2b})
correspond to the parts of the pressure generated by  the sources in
each body separately.

It is interesting to see how the global equilibrium is restored when
$T_1\rightarrow T_2=T$ in Eq.~(\ref{corr2b}). In this case
Eq.~(\ref{corr2b}) can be written as
\begin{eqnarray}
    \lefteqn{\!\!\!\!\!\!\!\left\langle E_\alpha[\omega;\textbf{r}]E_\beta^{\dag}[\omega';\textbf{r}']\right\rangle_{\textrm{sym}}=
    \frac{\hbar}{2}\coth\left(\frac{\hbar\omega}{2k_B
T}\right)\delta\left(\omega-\omega'\right)\times}\notag\\
   && \!\!\!\!\int_{V_1+V_2}d{\bf r}''\varepsilon''(\omega,{\bf r}'')G_{\alpha\gamma}[\omega;{\bf r},{\bf r}'']
    G_{\gamma\beta}^*[\omega;{\bf r}'',{\bf r}'].\label{correq}
\end{eqnarray}
The integral over the product of two Green functions is connected
with the imaginary part of the single Green function by the
important \cite{Eckhardt,Kazarinov} relation
\begin{eqnarray}
\lefteqn{\!\!\!\!\!\!\!\!\!\!\!\!\!\!\!\int_{\Omega}d{\bf r}''\varepsilon''(\omega,{\bf r}'')G_{\alpha\gamma}[\omega;{\bf r},{\bf r}'']
    G_{\gamma\beta}^*[\omega;{\bf r}'',{\bf r}']=}\notag\\
 && \;\;\;\;\;\; \;\;\;\;\;\;\;\;\;\;\; \;\;\;\;\;\;\;\;\;\;\; \;\;\;\;\;  4\pi\textrm{Im}G_{\alpha\beta}[\omega;{\bf r},{\bf r}'],
    \label{magic}
\end{eqnarray}
where $\Omega$ is a volume restricted by a surface where the Green
function vanishes. Keeping in mind that in the vacuum gap
$\varepsilon''=0$, one can extend the integration in
Eq.~(\ref{correq}) over the the whole space and using (\ref{magic})
one recovers the well-known form of the electric fields
fluctuation-dissipation theorem \cite{LLP} valid at a global thermal
equilibrium:
\begin{eqnarray}
   \lefteqn{\!\!\!\!\! \left\langle
    E_\alpha[\omega;\textbf{r}]E_\beta^{\dag}[\omega';\textbf{r}']\right\rangle_{\textrm{sym}}=}\notag\\
   && 2\pi\hbar\coth\left(\frac{\hbar\omega}{2k_B
T}\right)\textrm{Im}G_{\alpha\beta}[\omega;{\bf r},{\bf
r}']\delta\left(\omega-\omega'\right).
\label{correq2}
\end{eqnarray}
Notice that all fluctuations presented in this section include both
the vacuum ($T=0$) and the thermal fluctuations. These can be
identified  with the first and second terms, respectively, of the
r.h.s. of the identity
\begin{equation}
\coth\left(\frac{\hbar\omega}{2k_B
T}\right)= \textrm{sign}(\omega)\left(1+\frac{2}{e^{\hbar|\omega|/k_BT}-1}\right),\;\;\;\omega\neq0.
\label{T00wwetot}
\end{equation}
%
%If $\omega=0$ the imaginary part of the response functions in all previous equations vanishes.
%
%%%%%%%%%%%%%%%%%%%%%%%%%%%%%%%%%%%%%%%%%%%%%%%%%%%%%%%%%%%%%%%%%%%%%%%%%%%%%%%%%%%%%%%%%%%%%
\subsection{\label{PresGen}The pressure in terms of fluctuations}
%%%%%%%%%%%%%%%%%%%%%%%%%%%%%%%%%%%%%%%%%%%%%%%%%%%%%%%%%%%%%%%%%%%%%%%%%%%%%%%%%%%%%%%%%%%%%%
The pressure (\ref{force}) can be presented in terms of the Fourier transformed fields correlations:
\begin{eqnarray}
\lefteqn{\!\!\!\!\!\!\!\!\!\!\!\!\!\!\!\!P^{\textrm{neq}}(T_1,T_2,l)=-\frac{1}{8\pi}\iint\frac{\textrm{d}\omega}
{2\pi}\frac{\textrm{d}\omega'}{2\pi}e^{-i(\omega-\omega')t}\;\times} \notag\\
&&\;\;\;\;\;\;\;\;\Lambda_{\alpha\beta}
\left[\left\langle E_{\alpha}[\omega;{\bf
r}]E^{\dag}_{\beta}[\omega';{\bf r}']\right\rangle+\right.\notag\\
&&\left.\;\;\;\;\;\;\;\;\;\;\;\;\;\;\;\;\;\;\;\;
\left\langle B_{\alpha}[\omega;{\bf
r}]B^{\dag}_{\beta}[\omega';{\bf
r}']\right\rangle\right]\Big{|}_{{\bf r}={\bf r}'}.
\label{l;ivh}
\end{eqnarray}
Here the electric and magnetic contributions to the total pressure
are explicit, and ${\bf r}$ is a point inside of the vacuum gap. The stress tensor is in fact constant in the vacuum gap
due to the momentum conservation required by a stationary
configuration (see discussion in section \ref{sec:sfunctionnnn}). In
this equation we omitted the
symmetrization index since the average is taken at the same point ${\bf r}={\bf r}'$. Using Eq.(\ref{Max1}) it is useful to rewrite
expression (\ref{l;ivh}) in terms of the electric fields only
\cite{Matloob}  as
\begin{eqnarray}
\lefteqn{\!\!\!\!\!\!\!P^{\textrm{neq}}(T_1,T_2,l)=-\frac{1}{8\pi}\iint\frac{\textrm{d}\omega}{2\pi}\frac{\textrm{d}\omega'}{2\pi}e^{-i(\omega-\omega')t}\times}\notag\\
&&\;\;\;\;\;\;\;\;\;\;\;\;\;\;\;\;\;\;\;\;\Theta_{\delta\nu}\left[\left\langle E_{\delta}[\omega;{\bf r}]E^{\dag}_{\nu}[\omega';{\bf r}']\right\rangle\right]\Big{|}_{{\bf r}={\bf r}'}.
\label{l;ivhaa}
\end{eqnarray}
Here the pressure is expressed in terms of the correlations (\ref{corr2b}), and the operator
\begin{gather}
\Theta_{\delta\nu}=\Lambda_{\alpha\beta}\left(\delta_{\alpha\delta}\delta_{\beta\nu}+\frac{1}{k^2}
\epsilon_{\alpha\gamma\delta}\epsilon_{\beta\eta\nu}\partial_{\gamma}\partial'_{\eta}\right)
\label{pejce}
\end{gather}
selects the \textit{electric} and \textit{magnetic contributions}, given by the first and the second term in Eq.(\ref{pejce}), respectively.

From Eq.(\ref{T00wwetot}) it is possible to express the total pressure as the sum
\begin{gather}
P^{\textrm{neq}}(T_1,T_2,l)=P_{0}(l)+P_{\textrm{th}}^{\textrm{neq}}(T_1,T_2,l),
\label{ivjw}
\end{gather}
where the contribution of the zero-point ($T=0$) fluctuations,
$P_{0}(l)$, is separated from that produced by the thermal fluctuations,
$P_{\textrm{th}}^{\textrm{neq}}(T_1,T_2,l)$. Furthermore, thanks to equation (\ref{corr2b}) it is possible to express the
thermal component of the pressure acting between the bodies as the
sum of two terms
\begin{gather}
P_{\textrm{th}}^{\textrm{neq}}(T_1,T_2,l)= P_{\textrm{th}}^{\textrm{neq}}(T_1,0,l)+P_{\textrm{th}}^{\textrm{neq}}(0,T_2,l).
\label{òcjòoej}
\end{gather}
The pressure at thermal equilibrium $P^{\textrm{eq}}(T,l)$, being a particular case of (\ref{ivjw}), can be written  as
\begin{equation}
P^{\textrm{eq}}(T,l)=P_{0}(l)+P_{\textrm{th}}^{\textrm{eq}}(T,l).
\label{ivjwSummary}
\end{equation}
The pressures $P_{0}(l)$ and $P_{\textrm{th}}^{\textrm{eq}}(T,l)$ are  given by Eq.(\ref{l;ivhaa}), where the field fluctuations are provided by Eq.~(\ref{correq2}) after the substitution, respectively, of:
\begin{equation}
\coth\left(\frac{\hbar\omega}{2k_B
T}\right)\rightarrow \textrm{sign}(\omega),
\label{T00wwe}
\end{equation}
\begin{equation}
   \coth\left(\frac{\hbar\omega}{2k_B
T}\right)\rightarrow\frac{2\;\textrm{sign}(\omega)}{e^{\hbar|\omega|/k_BT}-1}.
\label{cot}
\end{equation}
If one simply perform such substitutions, it is well know that equation (\ref{l;ivhaa}) diverges at $T=0$, and  contains constant ($l$-independent) terms in the thermal part. The divergence has the same origin as the usual divergence of the zero-point fields energy in quantum electrodynamics, while the constant terms are related to the fact that we consider infinite bodies, and hence we neglect the pressure of the radiation exerted on the remote, external surfaces of the two bodies. To recover the exact finite value for the pressures $P_{0}(l)$, and exclude the constant terms in $P_{\textrm{th}}^{\textrm{eq}}(T,l)$, one should regularize the
Green function in the r.h.s of Eq.(\ref{correq2}) by subtracting the
bulk part $G_{ij}^{\textrm{bu.}}$, corresponding to a field
produced by a point-like dipole in an homogeneous and infinite
dielectric \cite{LP,TomasLif,Tomas95}. In fact, the Green function
with both the observation point ${\bf r}$ and the source point ${\bf
r}'$ in the vacuum gap  (see section \ref{sec:MMEss}) is given by the sum
\begin{equation}
G_{ij}\left[\omega;{\bf r},{\bf r}'\right]=G_{ij}^{\textrm{sc.}}\left[\omega;{\bf r},
{\bf r}'\right]+G_{ij}^{\textrm{bu.}}\left[\omega;{\bf r},{\bf r}'\right],
\label{llssy}
\end{equation}
of a \textit{scattered} and a \textit{bulk} term. The subtraction of the
bulk term corresponds to the subtraction of the pressure at
$l\rightarrow\infty$, as prescribed after Eq.(\ref{force}).
The expressions for the pressure at thermal equilibrium are given explicitly in section \ref{sec:LifTewwqq}.\\

Concerning the thermal pressure out of thermal equilibrium $P_{\textrm{th}}^{\textrm{neq}}(T_1,T_2,l)$  of Eq. (\ref{ivjw}), it can be obtained from Eq.(\ref{l;ivhaa}) by using (\ref{corr2b}) and the substitution (\ref{cot}).
Also in this case the thermal pressure $P_{\textrm{th}}^{\textrm{neq}}(T_1,T_2,l)$ contains an $l$-dependent and a constant term, as it happens for $P_{\textrm{th}}^{\textrm{eq}}(T,l)$ before being regularized. Differently from the equilibrium case, here the origin of the constant terms is not only due to the absence of the pressure acting on the remote surfaces,  but is also related to the fact that out of thermal equilibrium there is a net momentum transfer between the bodies. In this case the constant terms can remain also after considering bodies of finite thickness, and can even be different for the two bodies, depending on the external radiations. In the sections \ref{sec:LifTerNONEq} and \ref{CompPrev} we will calculate $P_{\textrm{th}}^{\textrm{neq}}(T_1,T_2,l)$ for two bodies filling two infinite half-spaces, and we will mainly discuss the pure $l$-dependent component. The constant terms will be discussed in section \ref{sec:freeedd} for the general case of bodies of finite thickness, with impinging the external radiations at different temperatures.
%
%%%%%%%%%%%%%%%%%%%%%%%%%%%%%%%%%%%%%%%%%%%%%%%%%%%
\subsection{\label{sec:eihvowiugh}Electromagnetic waves in surface optics}
%%%%%%%%%%%%%%%%%%%%%%%%%%%%%%%%%%%%%%%%%%%%%%%%%%
In this work we formulate the electromagnetic problem in terms of
$s$- and $p$-polarized vector waves and in terms of the Fresnel
coefficients for the interfaces \cite{Sipe}. Such notations are very
useful in surface optics. We will also employ the angular
spectrum representation for the description of the EM  and
polarization vectors.

If $\hat{\bf{x}}, \hat{\bf{y}}$ and $\hat{\bf{z}}$ are the
coordinate unit vectors (with real norm equal to $1$), one can write
the position vector as ${\bf r}={\bf R}+z\;\hat{\bf{z}}$, where
 the capital letter refers to vectors parallel to the interface
[${\bf R}\equiv(R_x,R_y,0)$]. Let us write the electromagnetic
(complex) wave vector in the medium $m$ with the complex dielectric
function
$\varepsilon_m(\omega)=\varepsilon_m'(\omega)+i\varepsilon_m''(\omega)$
as
\begin{equation}
{\bf q}^{(m)}(\pm)={\bf Q}\pm q_z^{(m)}\;\hat{\bf z}.
\label{12wv}
\end{equation}
Here the sign $(+)$ corresponds to an upward-propagating (or
evanescent) wave, and the sign $(-)$ corresponds to a
downward-propagating (or evanescent) wave. The vector ${\bf Q}\equiv
(Q_x,Q_y,0)$ is the projection (always real) of ${\bf q}^{(m)}(\pm)$
on the interface and the $z-$component of the wave vector, and 
\begin{gather}
q_z^{(m)}=\sqrt{\varepsilon_{m}\;k^2-Q^2}, 
\label{12qz}
\end{gather}
is a complex number with a positive imaginary part, and positive
real part in case $\textrm{Im}q_z^{(m)}=0$. Real and imaginary
parts of $q_z^{(m)}$ are expressed by the following relations:
\begin{eqnarray}
\textrm{Re}q_z^{(m)}=\sqrt{\frac{1}{2}\left[|\varepsilon_{m}(\omega)k^2-Q^2|+(\varepsilon_{m}'(\omega)k^2-Q^2)\right]},
\label{hhddhh}\\
\textrm{Im}q_z^{(m)}=\sqrt{\frac{1}{2}\left[|\varepsilon_{m}(\omega)k^2-Q^2|-(\varepsilon_{m}'(\omega)k^2-Q^2)\right]}.
\label{eocov}
\end{eqnarray}
Then, if the medium $m$ is non-absorbing
($\varepsilon_m''=0$), for $Q\leq\sqrt{\varepsilon_m'}\;k$ the wavevector
 $q_z^{(m)}$ is real and corresponds to a wave propagating in the
medium $m$, while for $Q>\sqrt{\varepsilon_m'}\;k$ the wavevector
$q_z^{(m)}$ is imaginary and corresponds to evanescent wave in the
medium $m$. The following identities will be useful:
\begin{eqnarray}
2\;\textrm{Im}\;q_z^{(m)}\;\textrm{Re}\;q_z^{(m)}&=&k^2\;\varepsilon_{m}''(\omega),
\label{idim}\\
\left(Q^2+|q_z^{(m)}|^2\right)\textrm{Re}\;q_z^{(m)}& =&k^2\;\textrm{Re}\left(\varepsilon_{m}^*(\omega)\;q_z^{(m)}\right),
\label{idim2}\\
\left(Q^2-|q_z^{(m)}|^2\right)\textrm{Im}\;q_z^{(m)}&=&k^2\;\textrm{Im}\left(\varepsilon_{m}^*(\omega)\;q_z^{(m)}\right).
\label{idim3}
\end{eqnarray}
It is worth noticing that the wave vectors ${\bf q}^{(m)}(\pm)$ lie in \textit{the plane of
incidence} spanned by $\hat{\bf Q}$ and $\hat{\bf z}$. Than one can
introduce the $s$- and $p$-unit complex polarization vectors
\begin{gather}
{\bf e}_s^{(m)}(\pm)=\hat{\bf Q} \wedge \hat{\bf z},
\label{es}\\
{\bf e}_p^{(m)}(\pm)={\bf e}_s^{(m)}(\pm)\wedge\hat{\bf q}^{(m)}(\pm)=
\frac{Q\hat{\bf z} \mp q_z^{(m)}\hat{\bf Q}}{\sqrt{\varepsilon_{m}(\omega)}\;k},
\label{1epd}
\end{gather}
that are vectors transversal and longitudinal to that plane,
respectively. Usually the polarization vector ${\bf e}_s^{(m)}(\pm)$
[${\bf e}_p^{(m)}(\pm)$] is  called transverse electric (TE)
[transverse magnetic (TM)] since it
corresponds to the electric [magnetic] field transverse to the plane
of incidence.

Our geometry consists of two half-spaces labeled with $m=1,2$
separated by a vacuum gap. Inside of the gap the wave vector ${\bf
q}$ and the polarization vectors ${\bf e}_{\mu}(\pm)$
are not labeled and are obtained, respectively, from the definitions
(\ref{12wv}), (\ref{12qz}) and (\ref{es}), (\ref{1epd}) by omitting
the apices $^{(m)}$, and setting $\varepsilon_m=1$.

Finally we can introduce the well known reflection and transmission Fresnel
coefficients for the vacuum gap-dielectric
interfaces, which for the $s$- and $p$-wave components are: 
\begin{eqnarray}
r^s_{m}=\frac{q_z-q_z^{(m)}}{q_z+q_z^{(m)}}&,&r^p_{m}=\frac{q_z\varepsilon_m-q_z^{(m)}}{q_z\varepsilon_m+q_z^{(m)}},
\label{rsrp}\\
t^s_{m}=\frac{2\,q_z^{(m)}}{q_z^{(m)}+q_z}&,&t^p_{m}=\frac{2\;\sqrt{\varepsilon_m(\omega)}\;q_z^{(m)}}{q_z^{(m)}+q_z\varepsilon_m(\omega)}.
\label{tstp}
\end{eqnarray}
In particular, the coefficients $r_{m}$ relate the radiation in the
vacuum gap impinging the interface $m$ and its part reflected
back into the vacuum gap. The coefficients $t_{m}$ relate the
radiation impinging the interface $m$ from the interior of the
dielectric $m$ and its part transmitted into the vacuum gap (see
Appendix \ref{sec:GTomas}).
%
%%%%%%%%%%%%%%%%%%%%%%%%%%%%%%%%%%%%%%%%%%%%%%%%%%%%%%%%%%%%%%%%%%%%%%%%%%%%%%%%%%%%%%%%%%%%
\section{\label{sec:LifTewwqq}Pressure at thermal equilibrium}
%%%%%%%%%%%%%%%%%%%%%%%%%%%%%%%%%%%%%%%%%%%%%%%%%%%%%%%%%%%%%%%%%%%%%%%%%%%%%%%%%%%%%%%%%%%
In this section we briefly recall the main results of the pressure in a system at thermal equilibrium. We present the thermal component of the pressure as the sum of PW and EW components, and in terms of real frequencies, which  will result useful for the rest of the discussion. The results we show for the pressure at equilibrium  are regularized [see discussion after Eq.(\ref{cot})].\\

The Lifshitz surface-surface pressure at thermal equilibrium can be
expressed in terms of real frequencies as
\begin{eqnarray}
P^{\textrm{eq}}(T,l)&=&-\frac{\hbar}{2\pi^2}\int_0^{\infty}\textrm{d}\omega\coth{\left(\frac{\hbar\omega}{2k_BT}\right)}\notag\\
&&\times \textrm{Re}\left[\int_0^{\infty}\textrm{d}Q\;Q\;q_z\;g(Q,\omega)\right],
\label{ujhv}
\end{eqnarray}
where
\begin{eqnarray}
g(Q,\omega)&=&\sum_{\mu=s,p}\frac{r^{\mu}_{1}r^{\mu}_{2}e^{2iq_z l}}{D_{\mu}}\notag\\
&=&\sum_{\mu=s,p}\left[(r^{\mu}_{1}r^{\mu}_{2})^{-1}e^{-2iq_z l}-1\right]^{-1}.
\label{jhvuwi}
\end{eqnarray}
In the previous equation the multiple reflections are described by the factor
\begin{equation}
D_{\mu}=1-r^{\mu}_{1}r^{\mu}_{2}e^{2iq_zl},
\label{ggd}
\end{equation}
and the reflection Fresnel coefficients $r^{\mu}_{m}$ for the
vacuum-dielectric interfaces are defined in Eq.(\ref{rsrp}).\\

By performing the Lifshitz rotation on the complex plane it is possible
to write Eq.(\ref{ujhv})  in terms of imaginary frequencies:
\begin{eqnarray}
P^{\textrm{eq}}(T,l)&=& \frac{k_B\;T}{16 \pi l^3}\int_0^{\infty}\textrm{d}x \;x^2\bigg[
\frac{(\varepsilon_{10}+1)(\varepsilon_{20}+1)}{(\varepsilon_{10}-1)(\varepsilon_{20}-1)}e^x-1\bigg]^{-1}
\notag\\
&+& \frac{k_B T}{\pi c^3}\sum_{n = 1}^{\infty}
  \xi^{3}_{n} \int_1^{\infty}\textrm{d}p\;p^2\;g(p,i\xi_n),
  \label{jjsystaSummary}
\end{eqnarray}
where $p=\sqrt{1+c^2Q^2/\xi_n^2}$. The dielectric functions that
enter to $g(p,i\xi_n)$ must be evaluated at imaginary frequencies
$\varepsilon_{1,2}=\varepsilon_{1,2}(i\xi_n)$, where $\xi_n=2\pi k_B
Tn/\hbar$. In the first term of (\ref{jjsystaSummary}) we have also
introduced the static values of the dielectric functions
$\varepsilon_{10}= \varepsilon_1(0)$ and
$\varepsilon_{20}=\varepsilon_2(0)$.

The pressure at thermal equilibrium includes contributions from
zero-point fluctuations $P_0(l)$ and from thermal fluctuations
$P_{\textrm{th}}^{\textrm{eq}}(T,l)$ as Eq. (\ref{ivjwSummary})
shows. $P_0(l)$ can be extracted from (\ref{ujhv}) with the
substitutes (\ref{T00wwe}) or from (\ref{jjsystaSummary}) as the
limit of continuous imaginary frequency.
The final result for the $T=0$ pressure is
\begin{equation}
P_0(l)=\frac{\hbar}{2\pi^2 c^3}\int_0^{\infty}\textrm{d}\xi\int_1^{\infty}\textrm{d}p\;p^2\xi^3\;g(p,i\xi).
\label{L00}
\end{equation}
The pressure $P_0(l)$ admits two important limits, i.e. the Van der
Waals-London and the Casimir-Polder behaviors, valid  at small and
large distances, respectively, in respect to the characteristic length
scale $\lambda_{opt}$  fixed by the absorption spectrum
of the bodies (typically is of the order of fraction of microns).

The behavior of the thermal component $P_{\textrm{th}}^{\textrm{eq}}(T,l)$ is 
related to a second length scale, i.e. the
thermal wavelength 
\begin{equation}
\lambda_T\equiv\frac{\hbar c}{k_BT}, 
\label{lam_T}
\end{equation}
which at room temperature is $\approx7.6\;\mu m$.

Then, the zero-point fluctuations dominate over the thermal
contribution at small distances $l\ll\lambda_T$. In this limit
behavior of the pressure is determined by the characteristic length
scale $\lambda_{opt}\ll\lambda_T$. In the interval $\lambda_{opt} \ll l \ll \lambda_T$
one enters the Casimir-Polder regime where the pressure decays like
$1/l^4$. For distances $l\ll \lambda_{opt}$ the force instead
exhibits the $1/l^3$  van der Waals-London dependence. The
possibility of identifying the  Casimir-Polder  regime depends
crucially on the value of the temperature. The temperature should be
in fact sufficiently low in order to guarantee the condition
$\lambda_T \gg \lambda_{opt}$.

\bigskip
The last part of this section focuses on the thermal component of the pressure, that will be often used along the rest of the paper. The pressure $P_{\textrm{th}}^{\textrm{eq}}(T,l)$ can be obtained from (\ref{ujhv}) by using (\ref{cot}). Since such a component of the pressure will
be compared with that out of thermal equilibrium, we show here
explicitly its expression for PW and EW contributions:
\begin{widetext}
\begin{eqnarray}
P_{\textrm{th}}^{\textrm{eq,PW}}(T,l)&=&-\frac{\hbar}{\pi^2}\int_0^{\infty}\textrm{d}
\omega\frac{1}{e^{\hbar\omega/k_BT}-1}\int_0^k\textrm{d}Q\;Q\;q_z\;\sum_{\mu=s,p}
\frac{\textrm{Re}\left(r^{\mu}_{1}r^{\mu}_{2}\;e^{2iq_zl}\right)-|r^{\mu}_{1}\;r^{\mu}_{2}|^2}{|D_{\mu}|^2},
\label{eqPW}\\
P_{\textrm{th}}^{\textrm{eq,EW}}(T,l)&=&\frac{\hbar}{\pi^2}\int_0^{\infty}\textrm{d}
\omega\frac{1}{e^{\hbar\omega/k_BT}-1}\int_k^{\infty}\textrm{d}Q\;Q\;\textrm{Im}q_z\;
e^{-2l\textrm{Im}q_z}\;\sum_{\mu=s,p}\frac{\textrm{Im}\left(r^{\mu}_{1}r^{\mu}_{2}\right)}
{|D_{\mu}|^2}.
\label{eqEW}
\end{eqnarray}
\end{widetext}
In particular at high temperatures, or equivalently at large distances defined by the condition
\begin{equation}
l\gg\lambda_T,
\label{lTTSummary}
\end{equation}
the leading contribution to the pressure  is given by the expression
 for the total force \cite{LP}
\begin{gather}
P_{\text{th}}^{\text{eq}}(T,l)=
\frac{k_B T}{16\pi l^3}%
\int_{0}^{\infty }\!\!\!\textrm{d}x\;x^{2}\;\left[ \frac{\varepsilon
_{10}+1}{\varepsilon _{10}-1}  \frac{ \varepsilon
_{20}+1}{\varepsilon _{20}-1}\;e^{x}-1\right] ^{-1}
\label{AsimTOTEQ}.
\end{gather}
It corresponds to the first term in Eq.(\ref{jjsystaSummary})
and is entirely due to the thermal fluctuations of the EM field.
In Ref.~\cite{LP} the asymptotic behavior (\ref{AsimTOTEQ}) has been
found after the contour rotation in the complex $\omega$-plane of
the EW term (\ref{eqEW}), that is partially canceled by the PW term
(\ref{eqPW}).

One can note that in this regime only the static value of the
dielectric functions is relevant. The pressure (\ref{AsimTOTEQ}) is
proportional to the temperature, is independent from the Planck
constant as well as from the velocity of light. We will call this
equation the Lifshitz limit. The pressure (\ref{AsimTOTEQ}) can be
obtained from the thermal free energy  $\mathcal{F}=\mathcal{E}-T\mathcal{S}$ of the electromagnetic  field
(per unit area) according to the thermodynamic identity  $P=-\left(
\partial \mathcal{F}/\partial l\right) _{T}$, where $\mathcal{E}$ and $\mathcal{S}$ are the thermal energy and entropy, respectively. It is interesting to
note that, differently from the free energy, \textit{the thermal energy} $\mathcal{E}$ \textit{decreases exponentially with} $l$, which means that the pressure
(\ref{AsimTOTEQ}) has pure entropic origin \cite{Revzen}.

It is important that at large separations only the $p$-polarization
contributes to the force (see for, example in
\cite{AntezzaPhDthesis}, the detailed derivation of the PW and EW
components). The reason is that for
low-frequencies the $s$-polarized field is nearly pure
magnetic, but the magnetic field penetrates freely into a non-magnetic material \cite{Sve05}.

In the limit $\varepsilon _{10},\varepsilon _{20}\rightarrow \infty
$ we find the force between two metals 
\begin{equation}
P_{\text{th}}^{\text{eq,met}}(T,l)= \frac{k_BT}{8\pi l^{3}}\;\zeta(3).
\end{equation}
Let us empathize that this result
was obtained for interaction between real metals
\cite{Bos00}.  For ''ideal mirrors'' considered by
Casimir, both polarizations of electromagnetic
fields are reflected. In this case there will be an additional factor 2 in
Eq.(\ref{AsimTOTEQ}) due to the contribution of the $s-$polarization. This ideal
case can be realized using \textit{superconducting} mirrors.

It is useful to note that the  surface-surface pressure
$P^{\textrm{eq}}(T,l)$ given by the Lifshitz result
(\ref{jjsystaSummary}) hides a non trivial cancellation between the
components of the pressure related to real and imaginary values of
the EM wavevectors, leading, respectively,  to the propagating
(PW) and  evanescent  (EW)  wave contributions \cite{articolo2,Intravaia}. This study deserves
careful investigation since  for a configuration out of thermal
equilibrium  such cancellations are no longer present, and the PW
and EW contribution will provide different asymptotic behaviors. The
new effect, as we will see, is particularly important if one of the two bodies is a
rarefied gas.
%
%%%%%%%%%%%%%%%%%%%%%%%%%%%%%%%%%%%%%%%%%%%%%%%%%%%%%%%%%%%%%%%%%%%%%%%%%%%%%%%%%%%%%%%%%%%%
\section{\label{sec:LifTerNONEq}Pressure out of thermal equilibrium between two infinite dielectric half-spaces}
%%%%%%%%%%%%%%%%%%%%%%%%%%%%%%%%%%%%%%%%%%%%%%%%%%%%%%%%%%%%%%%%%%%%%%%%%%%%%%%%%%%%%%%%%%%
As was discussed above [see Eqs. (\ref{corr2b}) and (\ref{òcjòoej})] each body contributes separately to the thermal pressure. In particular the pressure resulting from the thermal fluctuations
in the body $1$ is
\begin{eqnarray}
\lefteqn{\;\;P_{\textrm{th}}^{\textrm{neq}}(T,0,l)=}\notag\\
&&\!\!\!\!\!\!-\frac{\hbar}{16\pi^3}\int_0^{\infty}\textrm{d}\omega\frac{\varepsilon_1''
(\omega)}{e^{\hbar\omega/k_BT}-1}\textrm{Re}\left[\Theta_{\delta\nu}S_{\delta\nu}^{(1)}
\left[\omega;{\bf r}_1,{\bf r}_2\right]\right]\Big{|}_{{\bf
r}_1={\bf r}_2},\notag\\
 \label{LifForcesssLLss}
\end{eqnarray}
where ${\bf r}_1={\bf r}_2$ is a point in the vacuum gap and the function $S$ is defined in Eq.(\ref{Sdef}). In 
Eq.(\ref{LifForcesssLLss}) we used the parity properties
$\varepsilon''(\omega)=-\varepsilon''(-\omega)$ and
$S_{\delta\nu}\left[\omega;{\bf r}_1,{\bf
r}_2\right]=S_{\delta\nu}^*\left[-\omega;{\bf r}_1,{\bf r}_2\right]$
to restrict the range of integration to the positive
frequencies. It is evident that
$P_{\textrm{th}}^{\textrm{neq}}(0,T,l)$ can be expressed similarly to
Eq.(\ref{LifForcesssLLss}), but with
$\varepsilon_1''(\omega)\rightarrow\varepsilon_2''(\omega)$ and
$S_{\delta\nu}^{(1)}\rightarrow S_{\delta\nu}^{(2)}$.

Below  we specify the expressions of the tensors $S_{\delta\nu}^{(1)}$ and
$S_{\delta\nu}^{(2)}$ (section \ref{sec:sfunctionnnn}), calculate the electric and magnetic contributions to the pressure (section \ref{sec:PressureEM}), and finally provide the result for the total pressure in terms of PW and EW components (section \ref{sec:totptrss}). The total pressure will be rewritten in a different form in section \ref{CompPrev} by using a powerful expansion in multiple reflections. In the present and in the next section \ref{CompPrev} the pressure is calculated for two infinite bodies [see discussion at the end of section \ref{PresGen}].
%
%%%%%%%%%%%%%%%%%%%%%%%%%%%%%%%%%%%%%%%%%%%%%%%%%%%%%%%%%%%%%%%%%%%%%%%%%%%%%%%%%
\subsection{\label{sec:sfunctionnnn}The $S$ functions}
%%%%%%%%%%%%%%%%%%%%%%%%%%%%%%%%%%%%%%%%%%%%%%%%%%%%%%%%%%%%%%%%%%%%%%%%%%%%%%%%%
%
In this subsection we show the result for the tensors
$S_{\delta\nu}^{(1)}$ and $S_{\delta\nu}^{(2)}$ defined by
Eq.~(\ref{Sdef}). In terms of the lateral Fourier transforms
$s_{\delta\nu}^{(1)}[\omega;{\bf Q},z_1,z_2]$ and
$s_{\delta\nu}^{(2)}[\omega;{\bf Q},z_1,z_2]$ one has
\begin{gather}
S_{\delta\nu}[\omega;{\bf r}_1,{\bf r}_2]=\int\frac{\textrm{d}^2{\bf Q}}
{(2\pi)^2}\;e^{i{\bf Q}\cdot\left({\bf R}_1-{\bf R}_2\right)}\;
s_{\delta\nu}[\omega;{\bf Q},z_1,z_2].
\label{erjhv}
\end{gather}
By choosing the $x$-axis parallel to the vector ${\bf D}={\bf
R}_1-{\bf R}_2$ and defining $\phi$ as the angle between ${\bf Q}$
and ${\bf D}$ one gets that $Q_x=Q\cos\phi$, $Q_y=Q\sin\phi$ and the
polarization vectors become
\begin{eqnarray}
\!\!\!\!\!{\bf e}_s^{(m)}(\pm) &=&\left(|\sin\phi|,-\cos\phi\;\sin\phi/|\sin\phi|,0\right),
\label{esc}\\
\!\!\!\!\!{\bf e}_p^{(m)}(\pm)&=&\frac{1}{\sqrt{\varepsilon_{m}}k}\left(\mp q_z^{(m)}\cos\phi,\mp q_z^{(m)}\sin \phi, Q\right).
\label{eppc}
\end{eqnarray}
Here it is evident that
$|{\bf e}_{s}^{(m)}(\pm)|^2=1$ and
\begin{gather}
|{\bf e}_{p}^{(m)}(\pm)|^2=\frac{Q^2+|q_z^{(m)}|^2}{|\varepsilon_{m}|\;k^2}.
\label{jjii}
\end{gather}

After explicit calculation using the Green function given in
appendix \ref{sec:GTomas} we find for the $s_{\delta\nu}^{(1)}$ and
$s_{\delta\nu}^{(2)}$ functions the explicit expressions
\begin{widetext}
\begin{eqnarray}
s_{\delta\nu}^{(1)}[\omega;{\bf Q},z_1,z_2]&=&\frac{4\pi^2 k^2}
{\varepsilon_1''(\omega)}\frac{\textrm{Re}q_z^{(1)}}{|q_z^{(1)}|^2}
\;\sum_{\mu=s,p}\;\frac{|t_{1}^{\mu}|^2}{|D_{\mu}|^2}\;\;|{\bf e}_{\mu}^{(1)}(+)|^2\;\times\notag\\
&&\left[e_{\mu,\delta}(+)e_{\mu,\nu}^{*}(+)\;e^{i(q_zz_1-q_z^{*}z_2)}+
e_{\mu,\delta}(+)e_{\mu,\nu}^{*}(-)\;e^{i(q_zz_1+q_z^{*}z_2)}\;e^{-2iq_z^{*}l}\;r_{2}^{\mu *}+\right.\notag\\
\notag\\
&&\left.e_{\mu,\delta}(-)e_{\mu,\nu}^{*}(+)\;e^{-i(q_zz_1+q_z^{*}z_2)}\;e^{2iq_zl}\;r_{2}^{\mu }+
e_{\mu,\delta}(-)e_{\mu,\nu}^{*}(-)\;e^{-i(q_zz_1-q_z^{*}z_2)}\;e^{-4\textrm{Im}q_zl}\;|r_{2}^{\mu }|^2\right],
\label{;ro;elrgbj}\\
\notag\\
s_{\delta\nu}^{(2)}[\omega;{\bf Q},z_1,z_2]&=&\frac{4\pi^2 k^2}{\varepsilon_2''(\omega)}
\;\frac{\textrm{Re}q_z^{(2)}}{|q_z^{(2)}|^2}\;e^{-2l\textrm{Im}q_z}\;\sum_{\mu=s,p}
\;\frac{|t_{2}^{\mu}|^2}{|D_{\mu}|^2}\;\;|{\bf e}_{\mu}^{(2)}(-)|^2\;\times\notag\\
\notag\\
&&\left[e_{\mu,\delta}(-)e_{\mu,\nu}^{*}(-)\;e^{-i(q_zz_1-q_z^{*}z_2)}+
e_{\mu,\delta}(-)e_{\mu,\nu}^{*}(+)\;e^{-i(q_zz_1+q_z^{*}z_2)}\;r_{1}^{\mu *}+\right.\notag\\
\notag\\
&&\left.e_{\mu,\delta}(+)e_{\mu,\nu}^{*}(-)\;e^{i(q_zz_1+q_z^{*}z_2)}\;r_{1}^{\mu }+
e_{\mu,\delta}(+)e_{\mu,\nu}^{*}(+)\;e^{i(q_zz_1-q_z^{*}z_2)}\;|r_{1}^{\mu }|^2\right],
\label{;ro;elrgxwwbj}
\end{eqnarray}
\end{widetext}
where $D_{\mu}$ is defined in Eq.(\ref{ggd}).

It is worth noticing that in the non equilibrium but stationary regime the fields correlation
functions $s^{(1,2)}$ are not uniform in the vacuum cavity, while on the contrary  the
Maxwell stress tensor $T_{zz}$ (which is related to the momentum flux) has the same value in each point of the vacuum gap.
This is valid also at equilibrium, and is a direct consequence of the momentum conservation required by a stationary configuration. To show this property one can  set $z_1=z_2=z$
in Eq.(\ref{;ro;elrgbj}), where the dependence on $z$ appears only in
the exponential factors [the same would happen for Eq.(\ref{;ro;elrgxwwbj})] . Let us note that now the first and the last terms in such an expression are
proportional to $e^{2z\textrm{Im}q_z}$ and $e^{-2z\textrm{Im}q_z}$,
respectively, while the second and the third terms are proportional
to $e^{-2z\textrm{Re}q_z}$ and $e^{2z\textrm{Re}q_z}$, respectively.
As it will be clear in the next section \ref{sec:PressureEM} the first and the last terms will be responsible for the PW contribution to the pressure (for
which $\textrm{Im}q_z=0$), while the second and the third terms will
be responsible for the EW contribution (for which $\textrm{Re}q_z=0$).
It is then evident that the position $z$ disappears in the
Maxwell stress tensor.
%
%%%%%%%%%%%%%%%%%%%%%%%%%%%%%%%%%%%%%%%%%%%%%%%%%%%%%%%%%%%%%%%%%%%%%%%%%%%%%%%%%%%%%%%%%%%%
\subsection{\label{sec:PressureEM}Electric and magnetic contributions to the pressure}
%%%%%%%%%%%%%%%%%%%%%%%%%%%%%%%%%%%%%%%%%%%%%%%%%%%%%%%%%%%%%%%%%%%%%%%%%%%%%%%%%%%%%%%%%%%
%
The pure electric contribution to the pressure
$P_{\textrm{th}}^{\textrm{neq}}(T,0,l)$ is due to the first
term in Eq.~(\ref{pejce})
\begin{widetext}
\begin{eqnarray}
\Lambda_{\alpha\beta}\delta_{\alpha\delta}\delta_{\beta\nu}\;S_{\delta\nu}^{(1)}[\omega;{\bf r}_1,{\bf r}_2]\;\Big{|}_{{\bf r}_1={\bf r}_2, z_1=0}&=&\left[S_{11}^{(1)}+S_{22}^{(1)}-S_{33}^{(1)}\right]\Big{|}_{{\bf r}_1={\bf r}_2, z_1=0}=
\frac{2\pi k^2}{\varepsilon_1''}\int_0^{\infty}\textrm{d}Q\;Q\frac{\textrm{Re}\;q_z^{(1)}}{|q_z^{(1)}|^2}\;\times\notag\\
&&\left[\frac{|t_{1}^s|^2}{|D_{s}|^2}\;|{\bf e}_{s}^{(1)}(+) |^2\left( 1+r_{2}^{s*}\;e^{-2iq_z^*l}+r_{2}^{s}\;e^{2iq_zl}+|r_{2}^{s}|^2\;e^{-4l\textrm{Im}q_z}\right)+\right.\notag\\
&&\left.\frac{|t_{1}^p|^2}{|D_{p}|^2}\;|{\bf e}_{p}^{(1)}(+) |^2\left( \frac{|q_z|^2-Q^2}{k^2}-\frac{|q_z|^2+Q^2}{k^2}\;r_{2}^{p*}\;e^{-2iq_z^*l}-\right.\right.\notag\\
&&\left.\left.\frac{|q_z|^2+Q^2}{k^2}\;r_{2}^{p}\;e^{2iq_zl}+\frac{|q_z|^2-Q^2}{k^2}\;|r_{2}^{p}|^2\;e^{-4l\textrm{Im}q_z}\right) \right],
\label{ookiljcoj}
\end{eqnarray}
\end{widetext}
while the magnetic contribution is related by the second term in
Eq.~(\ref{pejce}), and is given by
\begin{widetext}
\begin{eqnarray}
\frac{1}{k^2}\Lambda_{\alpha\beta}\epsilon_{\alpha\gamma\delta}\epsilon_{\beta\eta\nu}\partial_{\gamma}\partial'_{\eta}\;S_{\delta\nu}^{(1)}[\omega;{\bf r}_1,{\bf r}_2]\;\Big{|}_{{\bf r}_1={\bf r}_2, z_1=0}&=&\frac{1}{k^2}\int_0^{\infty}\frac{\textrm{d}Q}{2\pi}\;Q\int_0^{2\pi}\frac{\textrm{d}\phi}{2\pi}\;e^{iQDcos\phi}\left\{\partial_3\partial_3'(s_{11}+s_{22})+\right.\notag\\
&&\left.\left[Q^2(s_{33}-s_{22}-s_{11})+Q_x^2s_{11}+Q_y^2s_{22}+Q_xQ_y(s_{12}+s_{21})\right]+\right.\notag\\
&&\left.\left[i\partial_3(Q_ys_{23}+Q_xs_{13})-i\partial_3'(Q_ys_{32}+Q_xs_{31})\right]\right\}\Big{|}_{D=0, z_1=z_2=0},
\label{ookddk}
\end{eqnarray}
\end{widetext}
where $s=s^{(1)}$. One can show that, as it happens for the equilibrium case,  the magnetic contribution (\ref{ookddk})
coincides with the electric one (\ref{ookiljcoj}), after the interchange of the
polarization indexes
$s\leftrightarrow p$.  

%
%%%%%%%%%%%%%%%%%%%%%%%%%%%%%%%%%%%%%%%%%%%%%%%%%%%%%%%%%%%%%%%%%%%%%%%%%%%%%%%%%%%%%%%%%%%%
\subsection{\label{sec:totptrss}Final expression for the pressure}
%%%%%%%%%%%%%%%%%%%%%%%%%%%%%%%%%%%%%%%%%%%%%%%%%%%%%%%%%%%%%%%%%%%%%%%%%%%%%%%%%%%%%%%%%%%
%
Taking the sum of 
(\ref{ookiljcoj}) and (\ref{ookddk}) one finds that the pressure
$P_{\textrm{th}}^{\textrm{neq}}(T,0,l)$ in
Eq.~(\ref{LifForcesssLLss}) is
\begin{widetext}
\begin{eqnarray}
P_{\textrm{th}}^{\textrm{neq}}(T,0,l)&=&-\frac{\hbar}{8\pi^2}\int_0^{\infty}\textrm{d}\omega\frac{1}{e^{\hbar\omega/k_BT}-1}\int_0^{\infty}\textrm{d}Q\;Q\;\frac{\textrm{Re}q_z^{(1)}}{|q_z^{(1)}|^2}\notag\\
&&\times\left\{\frac{|t_{1}^{s}|^2}{|D_{s}|^2}\left[\left(q_z^{2}+|q_z|^2\right)\left(1+|r^s_{2}|^2e^{-4\textrm{Im}q_zl}\right)+2\left(q_z^{2}-|q_z|^2\right)\textrm{Re}\left(r^s_{2}e^{2iq_zl}\right)\right]+\right.\notag\\
&&\left.\frac{|t_{1}^{p}|^2}{|D_{p}|^2}\frac{Q^2+|q_z^{(1)}|^2}{|\varepsilon_1(\omega)|k^2}\left[\left(q_z^{2}+|q_z|^2\right)\left(1+|r^p_{2}|^2e^{-4\textrm{Im}q_zl}\right)+2\left(q_z^{2}-|q_z|^2\right)\textrm{Re}\left(r^p_{2}e^{2iq_zl}\right)\right]\right\}.
\label{newforce}
\end{eqnarray}
\end{widetext}
From this general expression one can extract the contribution of the
propagating waves (PW) in the empty gap, for which $q_z$ is real and
hence $q_z^{2}=|q_z|^2$, and the contribution of the evanescent
waves (EW), for which  $q_z$ is pure imaginary and hence
$q_z^{2}=-|q_z|^2$:
\begin{widetext}
\begin{eqnarray}
\!\!\!\!\!\!\!P_{\textrm{th}}^{\textrm{neq,PW}}(T,0,l)\!&=&\!\!-\frac{\hbar}{4\pi^2}\!\!\int_0^{\infty}\!\!\!\!\!\!\textrm{d}\omega\frac{1}{e^{\hbar\omega/k_BT}-1}\!\int_0^{k}\!\!\!\!\textrm{d}QQ\frac{\textrm{Re}q_z^{(1)}}{|q_z^{(1)}|^2}q_z^{2}\;\left[\!\frac{|t_{1}^{s}|^2}{|D_{s}|^2}\left(1+|r^s_{2}|^2\right)+ \frac{|t_{1}^{p}|^2}{|D_{p}|^2} \frac{Q^2+|q_z^{(1)}|^2}{|\varepsilon_1(\omega)|k^2} \left(1+|r^p_{2}|^2\right)\!\right],
\label{newforcePW}\\
\notag\\
\!\!\!\!\!\!\!P_{\textrm{th}}^{\textrm{neq,EW}}(T,0,l)\!&=&\!\!-\frac{\hbar}{2\pi^2}\!\!\int_0^{\infty}\!\!\!\!\!\!\textrm{d}\omega\frac{1}{e^{\hbar\omega/k_BT}-1}\!\int_k^{\infty}\!\!\!\!\!\!\textrm{d}QQ\frac{\textrm{Re}q_z^{(1)}}{|q_z^{(1)}|^2}q_z^{2}e^{-2l\textrm{Im}q_z} \left[\!\frac{|t_{1}^{s}|^2}{|D_{s}|^2}\;\textrm{Re}\left(r^s_{2}\right)+\frac{|t_{1}^{p}|^2}{|D_{p}|^2}\frac{Q^2+|q_z^{(1)}|^2}{|\varepsilon_1(\omega)|k^2}\;\textrm{Re}\left(r^p_{2}\right)\!\right].
\label{newforceEW}
\end{eqnarray}
\end{widetext}
Now, using helpful identities \cite{Henkelprivcomm}
\begin{gather}
\frac{\textrm{Re}q_z^{(1)}\;|t^s_{1}|^2}{|q_z^{(1)}|^2}
=\frac{\textrm{Re}q_z\left(1-|r^s_{1}|^2\right)+2\;\textrm{Im}q_z\;\textrm{Im}r^s_{1}}{|q_z|^2},
\label{elelelejj}\\
\frac{\textrm{Re}\left(\varepsilon_1^*(\omega)\;q_z^{(1)}\right)\;|t^p_{1}|^2}{|\varepsilon_1(\omega)|\;|q_z^{(1)}|^2}
=
\frac{\textrm{Re}q_z\left(1-|r^p_{1}|^2\right)+2\;\textrm{Im}q_z\;\textrm{Im}r^p_{1}}{|q_z|^2},
\label{elelele}
\end{gather}
and similar ones for $1 \leftrightarrow 2$, it is possible to
express $P_{\textrm{th}}^{\textrm{neq,PW}}(T,0,l)$ and
$P_{\textrm{th}}^{\textrm{neq,EW}}(T,0,l)$ as
\begin{widetext}
\begin{eqnarray}
P_{\textrm{th}}^{\textrm{neq,PW}}(T,0,l)&=&-\frac{\hbar}{4\pi^2}\int_0^{\infty}\textrm{d}
\omega\frac{1}{e^{\hbar\omega/k_BT}-1}\int_0^k\textrm{d}Q\;Q\;q_z\;\sum_{\mu=s,p}
\frac{\left(1-|r^{\mu}_{1}|^2\right)\left(1+|r^{\mu}_{2}|^2\right)}{|D_{\mu}|^2},
\label{neqPW}\\
P_{\textrm{th}}^{\textrm{neq,EW}}(T,0,l)&=&\frac{\hbar}{\pi^2}\int_0^{\infty}\textrm{d}
\omega\frac{1}{e^{\hbar\omega/k_BT}-1}\int_k^{\infty}\textrm{d}Q\;Q\; \textrm{Im}q_z\;
e^{-2l\textrm{Im}q_z}\;\sum_{\mu=s,p}\frac{\textrm{Im}\left(r^{\mu}_{1}\right)\textrm{Re}
\left(r^{\mu}_{2}\right)}{|D_{\mu}|^2}.
\label{neqEW}
\end{eqnarray}
\end{widetext}

Note that the PW term (\ref{neqPW}) contains a distance independent
contribution that will be discussed in the next section.

The pressure $P_{\textrm{th}}^{\textrm{neq}}(0,T,l)$ can be obtained
following the same procedure but using the function $s_{ij}^{(2)}$
given by Eq.~(\ref{;ro;elrgxwwbj}). The result can be obtained
without calculation simply by the interchange
$r^{\mu}_{1}\leftrightarrow r^{\mu}_{2}$ in Eqs.~(\ref{neqPW}) and
(\ref{neqEW}).
%
%%%%%%%%%%%%%%%%%%%%%%%%%%%%%%%%%%%%%%%%%%%%%%%%%%%%%%%%%%%%%%%%%%%%%%%%%%%%%%%%%%%%%%%%%%%
\section{\label{CompPrev}Alternative expression for the pressure}
%%%%%%%%%%%%%%%%%%%%%%%%%%%%%%%%%%%%%%%%%%%%%%%%%%%%%%%%%%%%%%%%%%%%%%%%%%%%%%%%%
%
The thermal pressure between two bodies in a configuration out of thermal equilibrium was derived in the previous section, and expressed in terms of the Eqs. (\ref{neqPW}) and (\ref{neqEW}).
In this section we present an alternative expression for such a pressure, explicitly in terms of the pressure at thermal equilibrium. In section \ref{CompPrev1} we discuss the case of bodies made of identical materials $\varepsilon_1=\varepsilon_2$, in section \ref{CompPrev2} we discuss the general case of bodies made of different materials, and finally in section \ref{sec:freddedd} we show numerical results for the pressure between different bodies held at different temperatures.    
%%%%%%%%%%%%%%%%%%%%%%%%%%%%%%%%%%%%%%%%%%%%%%%%%%%%%%%%%%%%%%%%%%%%%%%%%%%%%%%%%%%%%%%%%%%
\subsection{\label{CompPrev1}Pressure between identical bodies}
%%%%%%%%%%%%%%%%%%%%%%%%%%%%%%%%%%%%%%%%%%%%%%%%%%%%%%%%%%%%%%%%%%%%%%%%%%%%%%%%%
%
In the case of two identical materials the pressure between bodies
can be found without any calculations using the following simple
consideration. Let the body 1 be at temperature $T$ and the body 2
be at $T=0$, then the thermal pressure will be $P^{\textrm{neq}}_{\textrm{th}}(T,0,l)$.
Because of the material identity the pressure will be the same if we
interchange the temperatures of the bodies:
$P^{\textrm{neq}}_{\textrm{th}}(T,0,l)=P^{\textrm{neq}}_{\textrm{th}}(0,T,l)$. In general we know from Eq.(\ref{òcjòoej}) that  the
thermal part of the pressure is given by the sum of two terms each of them corresponding to a
configuration where only one of the bodies is at non-zero
temperature, i.e. 
$P^{\textrm{neq}}_{\textrm{th}}(T_1,T_2,l)=P^{\textrm{neq}}_{\textrm{th}}(T_1,0,l)+P^{\textrm{neq}}_{\textrm{th}}(0,T_2,l)$. It is now evident that at equilibrium, where $T_1=T_2=T$, the
latter equation gives $P^{\textrm{neq}}_{\textrm{th}}(T,0,l)=P^{\textrm{eq}}_{\textrm{th}}(T,l)/2$ and
we find for the total pressure
\begin{gather}
P_{\textrm{th}}^{\textrm{neq}}(T_1,T_2,l)=\frac{P_{\textrm{th}}^{\textrm{eq}}(
T_1,l)}{2}+\frac{P_{\textrm{th}}^{\textrm{eq}}(T_2,l)}{2}.
\label{Dorof.}
\end{gather}
Therefore, the pressure between identical materials is expressed
only via the equilibrium pressures at $T_1$ and $T_2$. The same result
was obtained for the first time by Dorofeyev \cite{Dorofeyev1} by am explicit calculation of the pressure. It is interesting to note that the equation (\ref{Dorof.}) is valid not only for the plane-parallel geometry, but for any couple of identical bodies of any shape displaced in a symmetric configuration with respect to a plane.
%
%%%%%%%%%%%%%%%%%%%%%%%%%%%%%%%%%%%%%%%%%%%%%%%%%%%%%%%%%%%%%%%%%%%%%%%%%%%%%%%%%%%%%%%%%%%
\subsection{\label{CompPrev2}Pressure between different bodies}
%%%%%%%%%%%%%%%%%%%%%%%%%%%%%%%%%%%%%%%%%%%%%%%%%%%%%%%%%%%%%%%%%%%%%%%%%%%%%%%%%
%
It is convenient to present the general expression of the pressure in a form which  reduces to Eq.~(\ref{Dorof.}) in the case of identical
bodies. It can be done using Eq.~(\ref{òcjòoej}) where $P^{\textrm{neq}}_{\textrm{th}}(T,0,l)$ is given by (\ref{neqPW}) and (\ref{neqEW}), and $P^{\textrm{neq}}_{\textrm{th}}(0,T,l)$
 is obtained from $P^{\textrm{neq}}_{\textrm{th}}(T,0,l)$ after the interchange
$r^{\mu}_1\leftrightarrow r^{\mu}_2$. 

In $P^{\textrm{neq}}_{\textrm{th}}(T,0,l)$ we can separate symmetric
and antisymmetric  parts in respect to permutations of the bodies $1\leftrightarrow2$.
The factors sensitive to such a permutations in (\ref{neqPW}) and
(\ref{neqEW}) are, respectively,
\begin{eqnarray}
\lefteqn{\!\!\!\!\!\!\left(1-|r_1|^2\right)\left(1+|r_2|^2\right)=}\notag\\
&&\;\;\;\;\;\;\;\;\;\;\;\left(1-|r_1r_2|^2\right)+\left(|r_2|^2-|r_1|^2\right),
\label{saPW}\\
\notag\\
\lefteqn{\!\!\!\!\!\!\textrm{Im}(r_1)\textrm{Re}(r_2)=\frac{1}{2}\textrm{Im}(r_1
r_2)+}\notag\\
&&\;\;\;\;\;\;\;\;\;\;\;\frac{1}{2}\left[\textrm{Im}(r_1)\textrm{Re}(r_2)-\textrm{Re}(r_1)\textrm{Im}(r_2)\right],
\label{saEW}
\end{eqnarray}
where we omitted the index $\mu$.  The symmetric parts,
$\left(1-|r_1r_2|^2\right)$ for PW and $\textrm{Im}(r_1 r_2)/2$ for
EW, are responsible for the equilibrium term
$P_{\textrm{th}}^{\textrm{eq}}( T,l)/2$ in the non-equilibrium
pressure as Eq.~(\ref{Dorof.}) shows.  Concerning the EW terms,  if one takes the symmetric part
of (\ref{neqEW}), one obtains exactly $P_{\textrm{th}}^{\textrm{eq,EW}}(
T,l)/2$, where $P_{\textrm{th}}^{\textrm{eq,EW}}( T,l)$ coincides
with the equilibrium EW component (\ref{eqEW}). The analysis of the PW term is more delicate, in fact if one  takes the
symmetric part $\left(1-|r_1r_2|^2\right)$ of (\ref{neqPW}), one obtains $\overline{P}_{\textrm{th}}^{\textrm{eq,PW}}(T,l)/2$ where
\begin{eqnarray}\label{eqPWbar}
\overline{P}_{\textrm{th}}^{\textrm{eq,PW}}(T,l)&=&-\frac{\hbar}{2\pi^2}\int_0^{\infty}\frac{\textrm{d}
\omega}{e^{\hbar\omega/k_BT}-1}\times\notag\\
&&\int_0^k\textrm{d}QQ\;q_z\sum_{\mu=s,p}
\frac{1-|r^{\mu}_{1}\;r^{\mu}_{2}|^2}{|D_{\mu}|^2}.
\end{eqnarray}
The above equation is different from $P_{\textrm{th}}^{\textrm{eq,PW}}(T,l)$ given by (\ref{eqPW}).

The difference has a clear origin. In fact the pressure out of equilibrium, from which  (\ref{eqPWbar}) is derived,  is calculated for bodies occupying two infinite  half-spaces. On the contrary the equilibrium pressure $P_{\textrm{th}}^{\textrm{eq,PW}}(T,l)$ was obtained after proper regularization, and hence taking into account the pressure exerted on the external surfaces of bodies of finite thickness [see discussion after Eq.(\ref{cot})].  Then the difference between (\ref{eqPW}) and
(\ref{eqPWbar}) is just a constant:
\begin{equation}\label{Pbar}
    \overline{P}_{\textrm{th}}^{\textrm{eq,PW}}(T,l)=
    P_{\textrm{th}}^{\textrm{eq,PW}}(T,l)-\frac{4\sigma T^4}{3c},
\end{equation}
where $\sigma=\pi^2 k_B^4/60c^2\hbar^3$  is the Stefan-Boltzmann
constant. Using the following multiple-reflection expansion of the factor $|D_{\mu}|^{-2}$:
\begin{equation}
\frac{1}{|1-Re^{2iq_zl}|^2}=\frac{1}{1-|R|^2}\left(1+2\textrm{Re}
\sum_{n=1}^{\infty}R^ne^{2inq_zl}\right), \label{expfourPW}
\end{equation}
where $R=r_1^{\mu}r_2^{\mu}$, it is not difficult to show  explicitly that (\ref{eqPW}) and
(\ref{eqPWbar}) are related by (\ref{Pbar}). The constant term in (\ref{Pbar}) comes from the first term of this expansion.

Collecting together the symmetric and  antisymmetric parts we can finally
present the  non-equilibrium pressure in the following useful form:
\begin{widetext}
\begin{eqnarray}
P_{\textrm{th}}^{\textrm{neq,PW}}(T_1,T_2,l)&=&\frac{P_{\textrm{th}}^
{\textrm{eq,PW}}( T_1,l)}{2}+\frac{P_{\textrm{th}}^{\textrm{eq,PW}}(T_2,l)}{2}-
B(T_1,T_2)+ \Delta P_{\textrm{th}}^{\textrm{PW}}( T_1,l)-\Delta
P_{\textrm{th}}^{\textrm{PW}}(T_2,l),
\label{neqPW2}\\
\notag\\
P_{\textrm{th}}^{\textrm{neq,EW}}(T_1,T_2,l)&=&\frac{P_{\textrm{th}}^
{\textrm{eq,EW}}( T_1,l)}{2}+
\frac{P_{\textrm{th}}^{\textrm{eq,EW}}(T_2,l)}{2}+\Delta P_{\textrm{th}}^
{\textrm{EW}}( T_1,l)-\Delta P_{\textrm{th}}^{\textrm{EW}}(T_2,l).
\label{neqEW2}
\end{eqnarray}
\end{widetext}
This is one of the main result of this paper.
Here $B(T_1,T_2)=2\sigma \left(T_1^4+T_2^4\right)/3c$ is a
$l$-independent term, discussed in Eq.(\ref{Pbar}). The equilibrium pressures $P_{\textrm{th}}^
{\textrm{eq,PW}}( T,l)$  and $P_{\textrm{th}}^
{\textrm{eq,EW}}( T,l)$ are
defined by Eqs.(\ref{eqPW}) and (\ref{eqEW}) and do not contain $l$-independent terms. The expressions
$\Delta P_{\textrm{th}}^{\textrm{PW}}(T,l)$ and $\Delta
P_{\textrm{th}}^{\textrm{EW}}(T,l)$   are antisymmetric  respect to
the interchange of the bodies $1\leftrightarrow 2$ and are defined as:
\begin{widetext}
\begin{eqnarray}
\Delta P_{\textrm{th}}^{\textrm{PW}}(T,l)&=&-\frac{\hbar}{4\pi^2}
\int_0^{\infty}\textrm{d}\omega\frac{1}{e^{\hbar\omega/k_BT}-1}
\int_0^k\textrm{d}Q\;Q\;q_z\;\sum_{\mu=s,p}\frac{|r^{\mu}_{2}|^2-
|r^{\mu}_{1}|^2}{|D_{\mu}|^2},
\label{PWA}\\
\notag\\
\Delta P_{\textrm{th}}^{\textrm{EW}}(T,l)&=&\frac{\hbar}{2\pi^2}
\int_0^{\infty}\textrm{d}\omega\frac{1}{e^{\hbar\omega/k_BT}-1}
\int_k^{\infty}\textrm{d}Q\;Q\; \textrm{Im}q_z\;e^{-2l\textrm{Im}q_z}
\sum_{\mu=s,p}\frac{\textrm{Im}\left(r^{\mu}_{1}\right)\textrm{Re}
\left(r^{\mu}_{2}\right)-\textrm{Im}\left(r^{\mu}_{2}\right)\textrm{Re}
\left(r^{\mu}_{1}\right)}{|D_{\mu}|^2}.
\label{EWI}
\end{eqnarray}
\end{widetext}
Let us note that the EW term (\ref{EWI}) goes to $0$ for
$l\rightarrow\infty$ because evanescent fields decay at large
distances. However, the PW term (\ref{PWA}) contains a
$l$-independent component since in the non-equilibrium situation
there is momentum transfer between bodies. This $l$-independent
component can be directly extracted from (\ref{PWA}) using the
expansion (\ref{expfourPW}). This expansion shows explicitly the
contributions from multiple reflections. The distance independent
term corresponds to the first term in the expansion
(\ref{expfourPW}), and it is related with the radiation that pass the cavity only once, i.e. without being reflected. Finally it is possible to write
$\Delta P_{\textrm{th}}^{\textrm{PW}}(T,l)$ as the sum $\Delta
P_{\textrm{th}}^{\textrm{PW}}(T,l)=\Delta
P_{\textrm{th},a}^{\textrm{PW}}(T)+\Delta
P_{\textrm{th},b}^{\textrm{PW}}(T,l)$, where the constant and the pure $l$-dependent terms are respectively
\begin{widetext}
\begin{eqnarray}
\Delta P_{\textrm{th},a}^{\textrm{PW}}(T)&=&-\frac{\hbar }{4\pi ^{2}}\int_{0}^{%
\infty }\textrm{d}\omega\frac{1}{e^{\hbar \omega /k_BT}-1}
\int_{0}^{k}\textrm{d}Q\;Q\;q_z\;\sum_{\mu=s,p} \frac{\left| r^{\mu}_{2}\right|^{2}-\left|r^{\mu}_{1}\right| ^{2}}{%
1-\left| r^{\mu}_{1} r^{\mu}_{2}\right|^{2}} ,
\label{PWAConst}\\
\Delta P_{\textrm{th},b}^{\textrm{PW}}(T,l)&=&-\frac{\hbar }{2\pi ^{2}}\sum_{n=1}^{\infty}\textrm{Re}\left\{\int_{0}^{%
\infty }\textrm{d}\omega\frac{1}{e^{\hbar \omega /k_BT}-1}
\int_{0}^{k}\textrm{d}Q\;Q\;q_z\;\sum_{\mu=s,p} \frac{\left| r^{\mu}_{2}\right|^{2}-\left|r^{\mu}_{1}\right| ^{2}}{%
1-\left| r^{\mu}_{1} r^{\mu}_{2}\right|^{2}}\;(r^{\mu}_{1} r^{\mu}_{2})^n\;e^{2inq_zl}\right\}.
\label{PWA22}
\end{eqnarray}
\end{widetext}
At thermal equilibrium $T_1=T_2=T$ the sum of (\ref{neqPW2}) and
(\ref{neqEW2}) provides  the Lifshitz formula except for the term
$-4\sigma T^4/3c$, which is canceled due to the
pressure exerted on the remote external surfaces of the bodies, as explicitly shown in the next section. Out of thermal equilibrium, but
for identical bodies, $r^{\mu}_{1}=r^{\mu}_{2}$, the antisymmetric terms disappears: 
$\Delta P_{\textrm{th}}^{\textrm{PW}}(T,l)=\Delta
P_{\textrm{th}}^{\textrm{EW}}( T,l)=0$. In this case Eq.~(\ref{Dorof.}) is reproduced.\\

It is now clear that, due to the antisymmetric terms,  Eq.(\ref{Dorof.}) is not valid if the two
bodies are different. The problem of the interaction between two bodies with different
temperatures was previously considered by Dorofeyev \cite{Dorofeyev1} and Dorofeyev, Fuchs and Jersch \cite{Dorofeyev2}. The
authors used a different method, based on the generalized Kirchhoff's law
\cite{Rytov}. The general formalism of \cite{Dorofeyev1} agrees with our equations
(\ref{PWAConst}), (\ref{PWA22}). However, our results are in disagreement with the results
of \cite{Dorofeyev2}, where Eq.~(\ref{Dorof.}) was found
to be valid also for bodies of different materials. So that we argue that the results of the last paper were
based on some inconsistent derivation.
%
%%%%%%%%%%%%%%%%%%%%%%%%%%%%%%%%%%%%%%%%%%%%%%%%%%%%%%%%%%%%%%%%%%%%%%%%%%%%%%%%%
\subsection{\label{sec:freddedd}Numerical results for the pressure between two different bodies out of thermal equilibrium}
\begin{figure}[ptb]
\begin{center}
\includegraphics[width=0.45\textwidth]{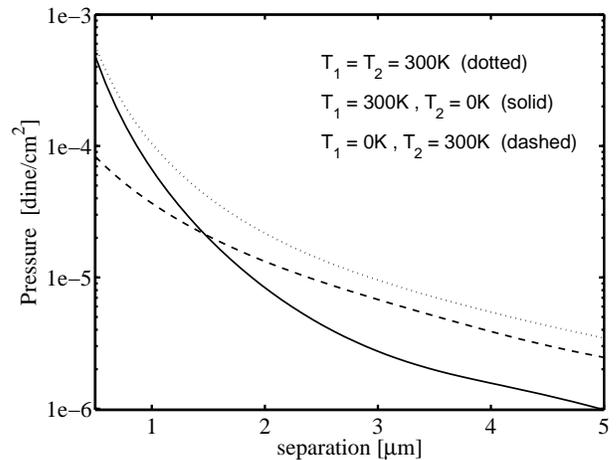}%{Pressure_th_FusSil_Silicon.eps}
\caption{Thermal component (only $l$-dependent part) of the pressure
out of equilibrium for fused silica-silicon system in the
configuration ($T_1=300$K, $T_2=0$K) [solid] and in the configuration
($T_1=0$K, $T_2=300$K) [dashed]. We plot also the thermal part of the
force at thermal equilibrium at $T=300$K [dotted].} \label{fig1FSS}
\end{center}
\end{figure}

\begin{figure}[ptb]
\begin{center}
\includegraphics[width=0.45\textwidth]{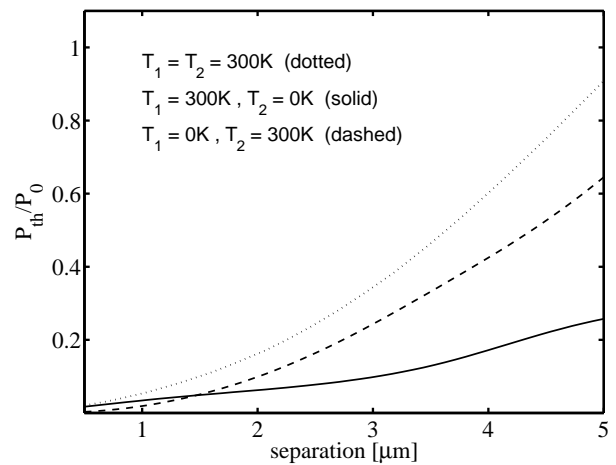}%{Pressure_Delta_th_and_Vac_FusSil_Silicon_PRL.eps}
\caption{Relative contribution of the thermal component of the
pressure  (only $l$-dependent part) out of equilibrium for fused
silica-silicon system in the configuration ($T_1=300$K, $T_2=0$K)
[solid], ($T_1=0$K, $T_2=300$K) [dashed], and
at thermal equilibrium at $T=300$K [dotted].} \label{fig2FSS}
\end{center}
\end{figure}

%%%%%%%%%%%%%%%%%%%%%%%%%%%%%%%%%%%%%%%%%%%%%%%%%%%%%%%%%%%%%%%%%%%%%%%%%%%%%%%%%
In this section we show the results of  the calculation of  the pressure between two different bodies, for configurations both in and out of thermal equilibrium.
In figures \ref{fig1FSS} and \ref{fig2FSS} we show the numerical results of the pressure for a system
made of fused silica (SiO$_2$) for the left-side body $1$ and low
conductivity silicon (Si) for the right-side body $2$. In both cases the
experimental values of the dielectric functions in a wide range of frequencies were taken from the
handbook \cite{Tropf}. In particular in Fig.\ref{fig1FSS} we show the thermal pressure
$P_{\textrm{th}}^{\textrm{neq}}(T_1,T_2,l)$, sum of
Eqs.(\ref{neqPW2}) and (\ref{neqEW2}), as a function of the separation $l$ between
$0.5\;\mu$m and $5\;\mu$m. Here we omit the $l$-independent
terms. The pressure is presented for the configuration ($T_1=300$K,
$T_2=0$K) [solid line] and for the configuration ($T_1=0$K,
$T_2=300$K) [dashed]. We plot also the thermal part of the force
at thermal equilibrium,  which is the sum of Eqs.(\ref{eqPW}) and
(\ref{eqEW}), at the temperature $T=300$K [dotted]. The sum of
the two configurations out of thermal equilibrium provides the force
at thermal equilibrium. In figure \ref{fig2FSS} we show the
relative contribution $P_{\textrm{th}}/P_0$ of the thermal component
(only the $l$-dependent terms) of the pressure
with respect to the vacuum pressure $P_0(l)$ given by Eq.(\ref{L00}).

\begin{figure}[ptb]
\begin{center}
\includegraphics[width=0.45\textwidth]{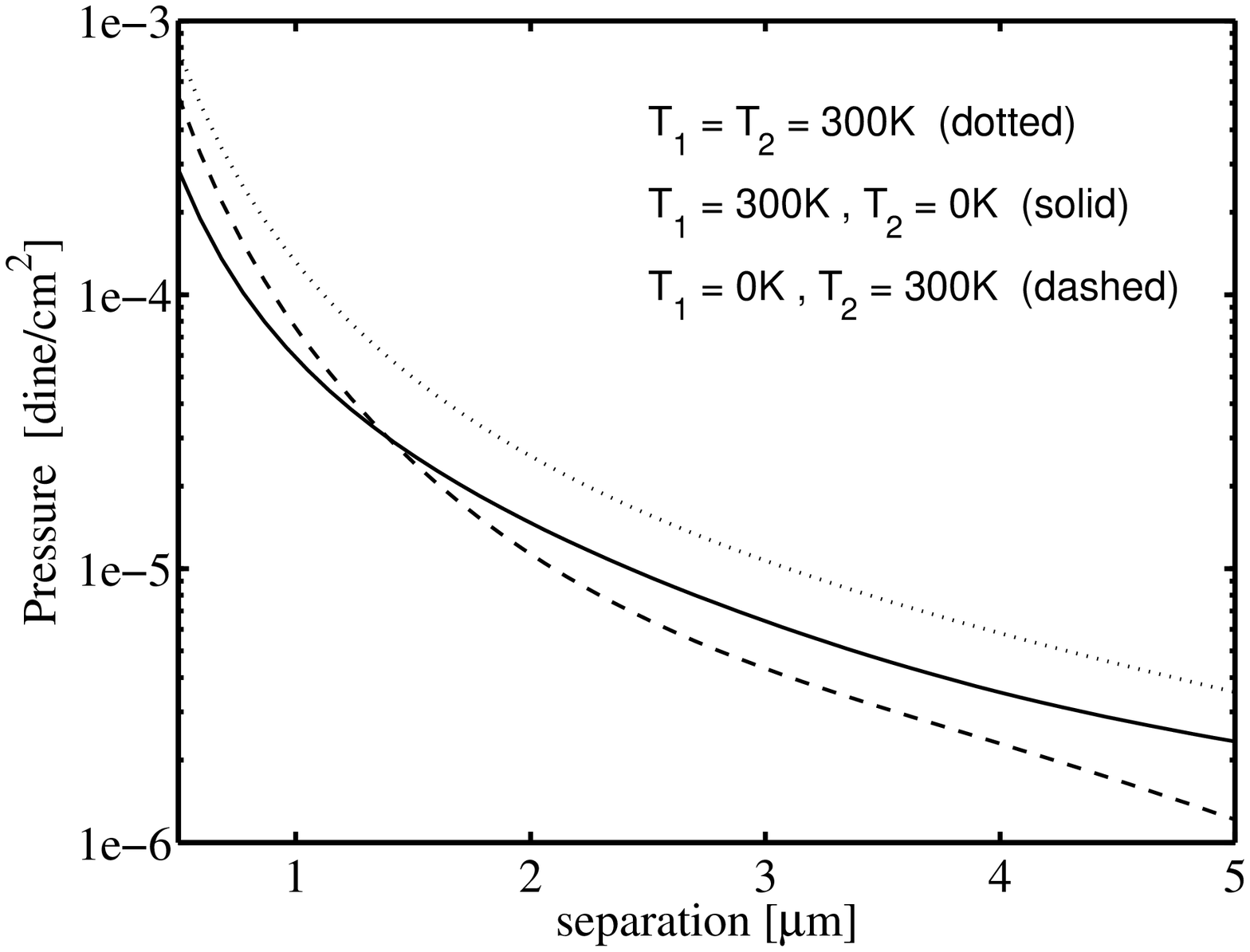}%{Pressure_th_sapphire_FusSil.eps}
\caption{Same of Fig.\ref{fig1FSS}, for the sapphire-fused silica system.} \label{fig1SFS}
\end{center}
\end{figure}

\begin{figure}[ptb]
\begin{center}
\includegraphics[width=0.45\textwidth]{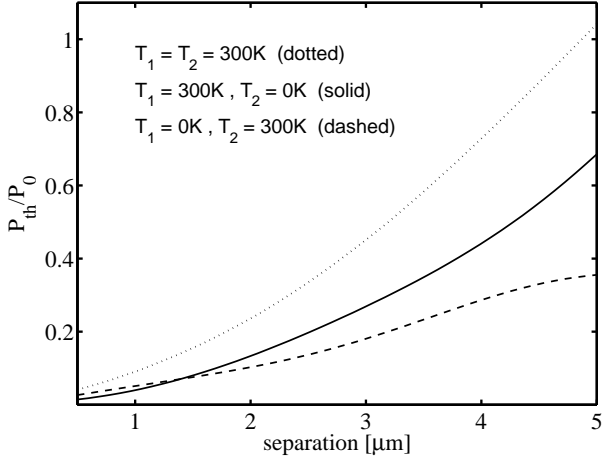}%{Pressure_Delta_th_and_Vac_sapphire_FusSil.eps}
\caption{Same of Fig.\ref{fig2FSS}, for the sapphire-fused silica system.} \label{fig2SFS}
\end{center}
\end{figure}

We performed the same analysis  for a different couple of materials, and in particular we considered  
sapphire (Al$_2$O$_3$) for the left-side body $1$, and 
fused silica (SiO$_2$) for the right-side body $2$. Also in this case the
experimental values of the dielectric functions were taken from the
handbook \cite{Tropf}. The results of such calculations are shown in figures \ref{fig1SFS} and \ref{fig2SFS}, where the same quantities of figure \ref{fig1FSS} and \ref{fig2FSS} were plotted.

From figures \ref{fig1FSS} and \ref{fig2FSS} it is evident that at
small separations the  pressure at ($T_1=300$K, $T_2=0$K) is lower
than  that at ($T_1=0$K, $T_2=300$K), and the situation  is inverted
at large separations. This is a characteristic feature of the
materials we use. In fact for the sapphire-fused silica system we
found the opposite behavior, as it is evident from figures \ref{fig1SFS} and \ref{fig2SFS}. This behavior is the result of the interplay between the relevant frequencies in the problem, i.e. the thermal wavelength $\lambda_T$, the separation $l$, and the different positions of the resonances in the dielectric functions for the different couples of materials.

%===================================================================================================
%===================================================================================================
%
%===================================================================================================
%===================================================================================================
%
%===================================================================================================
%===================================================================================================
%===================================================================================================
%===================================================================================================
%
%
%%%%%%%%%%%%%%%%%%%%%%%%%%%%%%%%%%%%%%%%%%%%%%%%%%%%%%%%%%%%%%%%%%%%%%%%%%%%%%%%%
\section{\label{sec:freeedd} Pressure between two thick slabs}
%%%%%%%%%%%%%%%%%%%%%%%%%%%%%%%%%%%%%%%%%%%%%%%%%%%%%%%%%%%%%%%%%%%%%%%%%%%%%%%%%
%
In sections \ref{sec:LifTerNONEq} and \ref{CompPrev} we derived and discussed the non-equilibrium pressure between two materials filling
two infinite half-spaces. We did not regularized the pressure, i.e. we did not considered the extra-pressure due to the presence of the external surfaces of the bodies. This would simply add new $l$-independent terms. We focused mainly on the $l$-dependent part. In this section we fill this gap, and derive the exact constant terms of the pressure for the general case of two bodies of finite thicknesses at different temperatures, in presence of external radiation.\\

At thermal equilibrium, due to the momentum's conservation theorem, the pressure cannot contain constant terms. In fact both Eq.(\ref{eqPW}) and (\ref{eqEW}) go to zero as $l$ goes to infinity. Here the regularization was performed by   subtracting the \textit{bulk part} of the full Green
function [see discussion after Eq.(\ref{cot})]. The inclusion of the bulk part
would add an extra $l$-independent term $-4\sigma T^4/3c$, as it is evident from the non regularized Eq.~(\ref{Pbar}). Physically the origin of this extra-term is due to the fact that the bodies are considered to be infinite, and hence, have no external surfaces. The presence of the external surfaces generates an extra pressure $4\sigma T^4/3c$, and finally the total pressure becomes $l$-independent. It is worth 
noticing that at thermal equilibrium the force acting on one body is
exactly the same (apart from the sign) of that acting on the second body.\\

\begin{figure}[ptb]
\begin{center}
\includegraphics[width=0.45\textwidth]{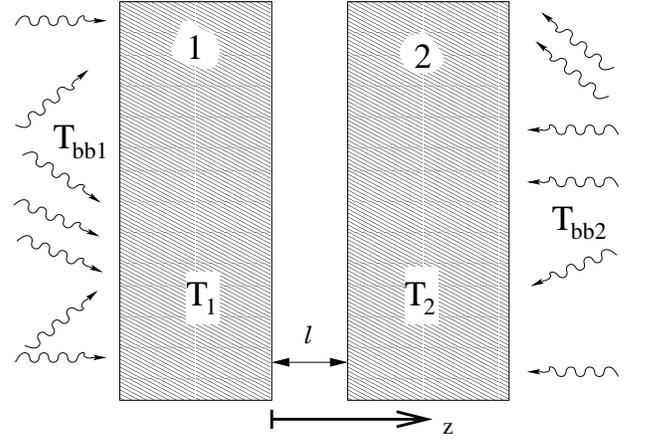}
\caption{Schematic figure of the two-slab system out of thermal equilibrium.}
\label{SlabFig}
\end{center}
\end{figure}

Out of thermal equilibrium, for bodies
occupying two half-spaces, one finds the non-regularized pressure given by the sum of Eqs.~
(\ref{neqPW}) and  (\ref{neqEW}). In this case the pressure contains  distance-independent
components, and is the same on both materials (apart from the sign). For bodies of finite thickness one should account for extra $l$-independent terms in the pressure due to the presence of two more interfaces between the bodies and the external regions (see Fig.\ref{SlabFig}) where, in general, the radiation is not in equilibrium with the bodies. In this configuration the pressure acting on the body $1$ can be different from that acting on the body $2$. It  should be noted that the new 
 $l-$independent terms should be added to
(\ref{neqPW2}) and  (\ref{neqEW2}), and originate from the PW waves only. Below we 
derive the result for such a general configuration, by manipulating Eq. (\ref{neqPW}).\\

Let us consider the case where both the bodies occupy
thick slabs, as represented in Fig.\ref{SlabFig}. 
On the left of the body $1$ impinges radiation at temperature $T_{bb1}$, while on the right of the body $2$
impinges radiation at temperature $T_{bb2}$. Then the pressure acting on
the body $1$  and body $2$ will be respectively:
\begin{eqnarray}
\!\!\!\!\!\!\!\!\!P_{1,\textrm{th}}^{\textrm{neq}}(T_{bb1},T_1,T_2,l)\!\!\!&=&\!\!\!
P_{\textrm{th}}^{\textrm{neq}}(T_1,T_2,l)+P_{\textrm{L}}(T_1,T_{bb1}),
\label{bo1}\\
\!\!\!\!\!\!\!\!\!P_{2,\textrm{th}}^{\textrm{neq}}(T_1,T_2,T_{bb2},l)\!\!\!&=&\!\!\!-P_{\textrm{th}}^{\textrm{neq}}(T_1,T_2,l)+P_{\textrm{R}}(T_2,T_{bb2}).
\label{bo2}
\end{eqnarray}
Here $P_{\textrm{th}}^{\textrm{neq}}(T_1,T_2,l)$ is the pressure out
of thermal equilibrium  given by the sum of (\ref{neqPW}) and
(\ref{neqEW}) for materials filling infinite half-spaces.
$P_{\textrm{L}}$ is the  pressure due to the presence of a new
left-side interface of the material $1$ while $P_{\textrm{R}}$ is
the pressure due to the presence of a new right-side interface of
the material $2$. Both $P_{\textrm{L}}$ and $P_{\textrm{R}}$ are
constant terms and include two contributions: the pressure of the external radiation impinging on the
outer interface and the back reaction produced by the emission of
radiation from the body to the vacuum half-space.

For thick enough slabs, it is possible to calculate the terms $P_{\textrm{L}}$ and $P_{\textrm{R}}$
using the expression of the pressure acting on a body $h$ which occupies an infinite half-space. In general it has a 
dielectric function $\varepsilon_h$, is at temperature $T_h$, and a thermal radiation with temperature $T_{bb}$ impinges on its free surface. There are two possible configurations. One corresponds to the body $h$ on the left
and radiation impinging from the right, the second correspond to the body on the right and radiation impinging from the left. In the two cases the pressures can be expressed in terms of the pressure between two infinite bodies $P_{\textrm{th}}^{\textrm{neq,PW}}(T_1,T_2,l)$ derived in the previous section, and are respectively:
\begin{eqnarray}
\lefteqn{\!\!\!\!\!\!P_{\textrm{R}}(T_h,T_{bb})=}\notag\\
&&\!\!\!\!\!\!\left[P_{\textrm{th}}^{\textrm{neq,PW}}(0,T_{bb},l)+P_{\textrm{th}}^{\textrm{neq,PW}}(T_h,0,l)\right]\Big{|}_{\varepsilon_2=1}^{\varepsilon_1\equiv\varepsilon_h},
\label{lchowssss}\\
\notag\\
\lefteqn{\!\!\!\!\!\!P_{\textrm{L}}(T_h,T_{bb})=}\notag\\
&&\!\!\!\!\!\!-\left[P_{\textrm{th}}^{\textrm{neq,PW}}(T_{bb},0,l)+P_{\textrm{th}}^{\textrm{neq,PW}}(0,T_h,l)\right]\Big{|}_{\varepsilon_1=1}^{\varepsilon_2\equiv\varepsilon_h}.
\label{lchows}
\end{eqnarray}
Here $P_{\textrm{th}}^{\textrm{neq,PW}}(T,0,l)$ and
$P_{\textrm{th}}^{\textrm{neq,PW}}(0,T,l)$ are given by
Eq.(\ref{neqPW}). After explicit calculations one find
\begin{eqnarray}
P_{\textrm{th}}^{\textrm{neq,PW}}(0,T_{bb},l)\Big{|}_{\varepsilon_2=1}^{\varepsilon_1\equiv\varepsilon_h}&=&
-\frac{2\sigma T_{bb}^4}{3c}-P_d(T_{bb}),
\label{wlvjwsssswe}\\
\notag\\
P_{\textrm{th}}^{\textrm{neq,PW}}(T_h,0,l)\Big{|}_{\varepsilon_2=1}^{\varepsilon_1\equiv\varepsilon_h}&=&-\frac{2\sigma T_{h}^4}{3c}+P_d(T_h),
\label{wlvjwswe}
\end{eqnarray}
where
\begin{equation}
P_d(T)=\frac{\hbar}{4\pi^2}\int_0^{\infty}\textrm{d}\omega\frac{1}{e^{\hbar\omega/k_BT}-1}\int_0^k\textrm{d}QQq_z\sum_{\mu=s,p}|r^{\mu}_{h}|^2,
\label{kkossj}
\end{equation}
and $r^{\mu}_{h}$ are defined similar to (\ref{rsrp}) but using the
dielectric function $\varepsilon_h$. Finally, we obtain the main result of this section, i.e. Eq.(\ref{lchowssss})
becomes
\begin{equation}
P_{\textrm{R}}(T_h,T_{bb})=-\frac{2\sigma (T_h^4+T_{bb}^4)}{3c}+P_d(T_h)-P_d(T_{bb}).
\label{RR}
\end{equation}
In the same way it is possible to calculate
$P_{\textrm{L}}(T_h,T_{bb})$ from Eq.(\ref{lchows}), and it is
evident that the result will be
\begin{equation}
P_{\textrm{L}}(T_h,T_{bb})=-P_{\textrm{R}}(T_h,T_{bb}).
\label{LL}
\end{equation}
At equilibrium $T_h=T_{bb}=T$ we find that
$P_{\textrm{R}}(T,T)=-P_{\textrm{L}}(T,T)=-4\sigma T^4/3c$ does not
depend on material characteristics and coincides with the pressure
of the black body radiation. It is also interesting to see that for
a white-body (W), corresponding to $|r^{\mu}_{h}|^2=1$, and for a
black-body (B),  corresponding to $|r^{\mu}_{h}|^2=0$, one obtains
\begin{eqnarray}
P_{\textrm{R}}(0,T)_{\textrm{W}}=-\frac{4\sigma T^4}{3c},&&P_{\textrm{R}}(0,T)_{\textrm{B}}=-\frac{2\sigma T^4}{3c},\label{P0T}\\
\notag\\
P_{\textrm{R}}(T,0)_{\textrm{W}}=0,&&P_{\textrm{R}}(T,0)_{\textrm{B}}=-\frac{2\sigma T^4}{3c},\label{PT0}.
\end{eqnarray}
From these relations one can see that
$P_{\textrm{R}}(0,T)_{\textrm{W}}/P_{\textrm{R}}(0,T)_{\textrm{B}}=2$, as it should be for the radiation pressure. Furthermore one has that that
$P_{\textrm{R}}(T,0)_{\textrm{W}}=0$. This is the consequence of the fact that if $|r^{\mu}_{h}|^2=1$
the radiation impinging on the surface from the interior of the
material is fully reflected and there is no flux of momentum outside
the body.

In the particular case when the external radiation is at equilibrium with the corresponding body, i.e. $T_{bb1}=T_1$ and $T_{bb2}=T_2$, from
(\ref{RR}) and (\ref{LL})  one obtains that Eqs. (\ref{bo1})
and (\ref{bo2}) become respectively
\begin{eqnarray}
\!\!\!\!\!\!\!\!P_{1,\textrm{th}}^{\textrm{neq}}(T_{bb1}=T_1,T_1,T_2,l)\!\!&=&\!\!P_{\textrm{th}}^{\textrm{neq}}(T_1,T_2,l)+\frac{4\sigma T_1^4}{3c},
\label{pivhefxx}\\
\!\!\!\!\!\!\!\!P_{2,\textrm{th}}^{\textrm{neq}}(T_1,T_2,T_{bb2}=T_2,l)\!\!&=&\!\!-P_{\textrm{th}}^{\textrm{neq}}(T_1,T_2,l)-\frac{4\sigma T_2^4}{3c}.
\label{pivhcsess}
\end{eqnarray}
If the whole system is at thermal equilibrium, $T_1=T_2=T$, the last
two equations give
\begin{eqnarray}
\lefteqn{\!\!\!\!\!\!\!\!P_{1,\textrm{th}}^{\textrm{neq}}(T,T,T,l)=-P_{2,\textrm{th}}^{\textrm{neq}}(T,T,T,l)=}\notag\\
&&P_{\textrm{th}}^{\textrm{neq}}(T,T,l)+\frac{4\sigma T^4}{3c}=P_{\textrm{th}}^{\textrm{eq}}(T,l).
\label{eqfindd}
\end{eqnarray}
This reproduces Eq.(\ref{Pbar}), where  $P_{\textrm{th}}^{\textrm{neq}}(T,T,l)\equiv\overline{P}_{\textrm{th}}^{\textrm{eq,PW}}(T,l)$.
%===================================================================================================
%===================================================================================================
%
%%%%%%%%%%%%%%%%%%%%%%%%%%%%%%%%%%%%%%%%%%%%%%%%%%%%%%%%%%%%%%%%%%%%%%%%%%%%%%%%%
\section{\label{sec:LongD}Long distance behavior of the surface-surface pressure}
%%%%%%%%%%%%%%%%%%%%%%%%%%%%%%%%%%%%%%%%%%%%%%%%%%%%%%%%%%%%%%%%%%%%%%%%%%%%%%%%%
Let us consider now the surface-surface pressure in the limit of large
separation. In this limit the relevant frequencies are $\omega\simeq c/l\ll
k_BT/\hbar$. If this frequency is smaller than the lowest absorption
resonance in the material, one can use the static
approximation for the dielectrics and change
$\varepsilon_i(\omega)\rightarrow \varepsilon_{0i}$. Some
dielectrics can have very low-lying resonances. For this case we
developed a special procedure that will be discussed later.

At thermal equilibrium the pressure is given by Eqs.(\ref{eqPW}) and
(\ref{eqEW}) for the PW and EW components, respectively. In the limit of large distances these components behave as 
\cite{AntezzaPhDthesis}
%
%\begin{widetext}
%
\begin{eqnarray}
\!\!\!\!\!\!\!\!P_{\textrm{th}}^{\textrm{eq,PW}}(T,l) &=&
\frac{k_BT\zeta(3)}{4\pi l^3},
\label{AsimPWEQ}\\
\!\!\!\!\!\!\!\!P_{\text{th}}^{\text{eq,EW}}(T,l) &=& -\frac{k_B
T\zeta(3)}{4\pi l^3}+\notag\\
&&\!\!\!\!\!\!\!\!\!\!\!\!\!\!\!\!\!\!\!\!\!\!\!\!\!\!\!\!\!\!\!\!\!\!\!\!\!\frac{k_B T}{16\pi l^3}\int_{0}^{\infty }\text{d}x\;x^{2}\left[
\frac{\varepsilon _{10}+1}{\varepsilon _{10}-1}  \frac{ \varepsilon
_{20}+1}{\varepsilon _{20}-1}\;e^{x}-1\right] ^{-1},
\label{AsimEWEQ}
\end{eqnarray}
%
%\end{widetext}
%
where $\zeta(3)\approx1.2021$ is the Riemann zeta-function. These
equations are both valid at the condition
\begin{equation}
l\gg\textrm{max}_{m=1,2}\left(\frac{\varepsilon_{m0}}{\sqrt{\varepsilon_{m0}-1}}\right)\lambda_T,
\label{LimPW}
\end{equation}
where $\lambda_T$ is defined in Eq.(\ref{lam_T}). The first term in
Eq.~(\ref{AsimEWEQ}) is canceled by the contribution from the
propagating waves (\ref{AsimPWEQ}), and their sum provides the well known result
for the total force at equilibrium (\ref{AsimTOTEQ}). It is worth noticing that the \textit{total} force at equilibrium is valid at the condition (\ref{lTTSummary}), which is significantly different from (\ref{LimPW}) if one of the two bodies is rarefied.

The surface-surface force in the non-equilibrium case is given by
Eqs.~(\ref{neqPW2}) and (\ref{neqEW2}). Omitting the $l$-independent
terms one finds for the large distance behavior the following
result \cite{ShortArticle}:
\begin{widetext}
\begin{eqnarray}
P_{\textrm{th}}^{\textrm{neq,PW}}(T,0,l)&=&\frac{k_BT\zeta(3)}
{16\pi l^3}\left[2-\frac{\sqrt{\varepsilon_{10}-1}-\sqrt{\varepsilon_{20}-1}}
{\sqrt{\varepsilon_{10}-1}+\sqrt{\varepsilon_{20}-1}}-\frac{\varepsilon_{20}
\sqrt{\varepsilon_{10}-1}-\varepsilon_{10}\sqrt{\varepsilon_{20}-1}}
{\varepsilon_{20}\sqrt{\varepsilon_{10}-1}+\varepsilon_{10}\sqrt{\varepsilon_{20}-1}}\right],
\label{AsimPW}\\
P_{\textrm{th}}^{\textrm{neq,EW}}(T,0,l)&=&\frac{k_BT}{8\pi^2 l^3}
\int_0^{\infty}\textrm{d}t\int_0^{\infty}\textrm{d}x\;\frac{x^2\;e^{-x}}{t}\;
\sum_{\mu=s,p}\;\frac{\textrm{Im}\left[r^{\mu}_{1}(t)\right]\textrm{Re}
\left[r^{\mu}_{2}(t)\right]}{|1-r^{\mu}_{1}(t)r^{\mu}_{2}(t)\;e^{-x}|^2}.
\label{AsimEW}
\end{eqnarray}
\end{widetext}
Here  $r^{\mu}_{m}(t)$ are the Fresnel
reflection coefficients (\ref{rsrp}) in
the static approximation $\varepsilon_{m}=\varepsilon_{m0}$, and $t$ is defined by the relation
$Q^2=k^2\left(1+t^2\right)$. Note that the equations (\ref{AsimPW})
and (\ref{AsimEW}) are also valid at the condition (\ref{LimPW}).

In the two following subsections \ref{sec:LongDPW} and \ref{sec:LongDEW} we will describe the procedure we used to calculate the large distance asymptotic behaviors (\ref{AsimPW}) and (\ref{AsimEW}) for the PW and EW components, respectively.
%
%%%%%%%%%%%%%%%%%%%%%%%%%%%%%%%%%%%%%%%%%%%%%%%%%%%%%%%%%%%%%%%%%%%%%%%%%%%%%%%%%
\subsection{\label{sec:LongDPW}Asymptotic behavior for PW}
%%%%%%%%%%%%%%%%%%%%%%%%%%%%%%%%%%%%%%%%%%%%%%%%%%%%%%%%%%%%%%%%%%%%%%%%%%%%%%%%%
In this subsection we derive the expansion of the PW contribution
$P_{\textrm{th}}^{\textrm{neq,PW}}(T,0,l)$ at large distances, just
anticipated in Eq.~(\ref{AsimPW}). We concentrate on the
$l$-dependent part only. One can start from Eq.~(\ref{neqPW}). It is
helpful to use the multiple-reflection expansion expressed by
Eq.~(\ref{expfourPW}). The first term in this expansion corresponds
to radiation which is emitted by one plate and absorbed by the other
one, without being reflected back. This is a distance independent
term which we omit. All the other terms of the sum give contribution
to the distance dependent part to which we are interested.

Let us introduce new variables and parameters in Eq.~(\ref{neqPW}),
i.e.
\begin{equation}
x=\frac{\hbar \omega }{k_BT},\; Q^{2}=k^2(1-t^2),\; \alpha
=\frac{\lambda_T}{2l}. \label{6}
\end{equation}
The limit of large distances corresponds to $\alpha \ll 1$. In terms
of these new variables one finds
\begin{widetext}
\begin{eqnarray}
P_{\textrm{th}}^{\textrm{neq,PW}}(T,0,l)=\frac{(k_BT)^4}{2\pi ^{2}\hbar^3
c^3} \sum_{n=1}^{\infty}\textrm{Re}\left\{\int_{0}^{ \infty
}\textrm{d}x\frac{x^3}{e^{x}-1}
\int_{0}^{1}\textrm{d}tt^2\sum_{\mu=s,p} \frac{\left(1-\left|
r^{\mu}_{1}\right|^{2}\right)\left(1+\left|r^{\mu}_{2}\right|
^{2}\right)}{ 1-\left| r^{\mu}_{1}
r^{\mu}_{2}\right|^{2}}\;\left(r^{\mu}_{1}
r^{\mu}_{2}\right)^n\;e^{intx/\alpha}\right\},
 \label{ppasde}
\end{eqnarray}
\end{widetext}
where the reflection coefficients as functions of $x$ and $t$ are
\begin{eqnarray}
r^{s}_{m}(t,x)&=&\frac{t-\sqrt{\varepsilon_m-1+t^{2}}}{t+\sqrt{\varepsilon_m-1+t^{2}}},\label{ft}\\
r^{p}_{m}(t,x)&=&\frac{\varepsilon_m t-\sqrt{\varepsilon_m
-1+t^{2}}}{\varepsilon_m t+\sqrt{\varepsilon_m-1+t^{2}}}, \label{8}
\end{eqnarray}
and  $\varepsilon_m=\varepsilon_m(k_BTx/\hbar)$ is a function of the
variable $x$. For $\alpha\ll1$ the integrand in (\ref{ppasde})
oscillates fast and it is possible to show that the relevant values
of variables in the integral are  $x\lesssim 1$ and
$t\sim\alpha/n$. Then, expanding the reflection coefficients for
small values of $t$ and integrating over $t$ explicitly one finds the
leading term in $\alpha$
%
%\begin{widetext}
%
\begin{eqnarray}
\lefteqn{\!\!\!\!\!\!P_{\textrm{th}}^{\textrm{neq,PW}}(T,0,l)=}\notag\\
&&\!\!\!\!\frac{k_BT}{8\pi
^{2}l^3}\sum_{n=1}^{\infty}\frac{1}{n^3}\int_{0}^{
\infty}\textrm{d}x\frac{\sin(n
x/\alpha)}{e^{x}-1}\sum_{\mu=s,p}g_{\mu}(x), \label{ppasdess}
\end{eqnarray}
%
%\end{widetext}
%
where the following functions of $x$ were introduced
\begin{equation}
g_{s}\left( x\right)=\frac{2\textrm{Re}\left( \beta_{1}\right) }{
\textrm{ Re}\left( \beta _{1}+\beta _{2}\right) },\;\;g_{p}\left(
x\right) =\frac{2\textrm{ Re}\left( \gamma _{1}\right) }{ \textrm{
Re}\left( \gamma _{1}+\gamma _{2}\right)}, \label{sisj}
\end{equation}
with
\begin{equation}
\beta_m(x) =\frac{1}{\sqrt{ \varepsilon_m-1}},\;\; \gamma_m(x)
=\frac{ \varepsilon_m }{\sqrt{\varepsilon_m -1}}. \label{10q}
\end{equation}
The leading contribution to
$P_{\textrm{th}}^{\textrm{neq,PW}}(T,0,l)$ comes from the region
$x\sim \alpha /n\ll 1$, where $e^{x}-1\approx x$. Note that one can
do this expansion only after explicit integration over $t$. After
the change of variable $y=nx/\alpha $ we obtain
\begin{eqnarray}
\lefteqn{\!\!\!\!\!\!P_{\textrm{th}}^{\textrm{neq,PW}}(T,0,l)=}\notag\\
&&\!\!\!\!\frac{k_BT}{8\pi
^{2}l^3}\sum_{n=1}^{\infty}\frac{1}{n^3}\int_{0}^{ \infty
}\textrm{d}y\frac{\sin y}{y}\sum_{\mu=s,p}g_{\mu}(\alpha y/n).
\label{ppasdede}
\end{eqnarray}
The relevant range of integration here is $y\sim1$, and then the
important frequencies in the dielectric functions entering in
Eq.(\ref{10q}) are of the order of $\hbar \omega \sim \alpha k_BT$.
Most of the dielectrics (but not all) at these frequencies have no
dispersion in the spectrum and one can take the static
approximation $g_{\mu}\left( \alpha y/n\right)\approx g_{\mu}\left(
0\right)$. In this case the integral in Eq.~(\ref{ppasdede}) can be
calculated explicitly:
\begin{equation}
P_{\textrm{th}}^{\textrm{neq,PW}}(T,0,l)= \frac{k_BT}{16\pi
l^3}\zeta(3)\left[g_s(0)+g_p(0)\right], \label{ppasdcc}
\end{equation}
where
\begin{eqnarray}
g_{s}\left( 0\right)&=&\frac{2\sqrt{\varepsilon_{20}-1}}{\sqrt{\varepsilon_{10}-1}+\sqrt{\varepsilon_{20}-1}},\\
g_{p}\left(0\right)&=&\frac{2\varepsilon_{10}\sqrt{\varepsilon_{20}-1}}{\varepsilon_{20}\sqrt{\varepsilon_{10}-1}+\varepsilon_{10}\sqrt{\varepsilon_{20}-1}}.
\label{sisjss}
\end{eqnarray}
Eq.~(\ref{ppasdcc}) coincides with (\ref{AsimPW}) after elementary
transformation. This expression is valid under the condition
(\ref{LimPW}) that justify the expansion on $t$ done for the
reflection coefficients (\ref{ft}) and (\ref{8}).

It is interesting to derive also the large distance behavior
(\ref{AsimPWEQ}) for the equilibrium case. To do this we can note
that the symmetric part of the non-equilibrium pressure in respect
to the interchange of the bodies coincides with one half of the
equilibrium pressure as Eq.~(\ref{neqPW2}) demonstrates. The
symmetric part of both $g_{s}(x)$ and $g_{p}(x)$ is equal to 1 and  we
immediately reproduce the result (\ref{AsimPWEQ}).
%
%%%%%%%%%%%%%%%%%%%%%%%%%%%%%%%%%%%%%%%%%%%%%%%%%%%%%%%%%%%%%%%%%%%%%%%%%%%%%%%%%
\subsection{\label{sec:LongDEW}Asymptotic behavior for EW}
%%%%%%%%%%%%%%%%%%%%%%%%%%%%%%%%%%%%%%%%%%%%%%%%%%%%%%%%%%%%%%%%%%%%%%%%%%%%%%%%%
In this subsection we show how to evaluate the asymptotic behavior of the EW
contribution to the pressure $P_{\textrm{th}}^{\textrm{neq,EW}}(T,0,l)$, whose result was anticipated in Eq.(\ref{AsimEW}). We start from the general expression for $P_{\textrm{th}}^{\textrm{neq,EW}}(T,0,l)$ given by Eq.~(\ref{neqEW}).  Substituting in this equation 
$x$ and $\alpha$ given by (\ref{6}), but defining $t$  as
$Q^2=k^2\left(1+t^2\right)$,  one finds for the pressure
\begin{eqnarray}
\lefteqn{\!\!\!\!\!\!\!\!\!\!\!\!P_{\textrm{th}}^{\textrm{neq,EW}}(T,0,l)=\frac{(k_BT)^4}{\pi^2\hbar^3c^3}
\int_0^{\infty}\textrm{d}x\frac{x^3}{e^x-1}\times}\notag\\
&&\!\!\!\!
\int_0^{\infty}\textrm{d}tt^2
e^{-xt/\alpha}\;\sum_{\mu=s,p}\frac{\textrm{Im}\left(r^{\mu}_{1}\right)\textrm{Re}
\left(r^{\mu}_{2}\right)}{|1-r^{\mu}_{1}r^{\mu}_{2}\;e^{-xt/\alpha}|^2}.
\label{newhyt}
\end{eqnarray}
Here the reflection coefficients are functions of $t$ and $x$ and take
the form
\begin{eqnarray}
r^{s}_{m}(t,x)&=&\frac{it-\sqrt{\varepsilon_m-1-t^{2}}}
{it+\sqrt{\varepsilon_m-1-t^{2}}},\label{fthy}\\
r^{p}_{m}(t,x)&=&\frac{i\varepsilon_m\; t-\sqrt{\varepsilon_m
-1-t^{2}}}{i\varepsilon_m\; t+\sqrt{\varepsilon_m-1-t^{2}}},
\label{8se}
\end{eqnarray}
with $\varepsilon_m=\varepsilon_m(k_BTx/\hbar)$.

Differently from the  PW component, here the relevant ranges of variables in
the integral (\ref{newhyt}) are $x\sim\alpha$ and $t\sim1$. Small values of
$t$ do not give significant contribution because the integrand is
suppressed by a factor $t$ coming from $\textrm{Im}\left(r^{\mu}_{1}\right)$,  that do not appears in the  PW
case.
Then for large distances it is possible to
expand on small values of $x$ and approximate $e^x-1\approx x$. It
is convenient to introduce the new variable $y=xt/\alpha$ instead of
$x$, for which the important range is now $y\sim1$. In terms of $y$ and
$t$ the pressure can be presented as
\begin{eqnarray}
\lefteqn{\!\!\!\!\!\!\!\!\!\!\!\!P_{\textrm{th}}^{\textrm{neq,EW}}(T,0,l)=\frac{k_B
T}{8\pi^2 l^3 }\int_0^{\infty} \frac{\textrm{d}t}{t}
\times}\notag\\
&&\;\;\;\;\;\int_0^{\infty}\textrm{d}yy^2
e^{-y}\sum_{\mu=s,p}\frac{\textrm{Im}\left(r^{\mu}_{1}\right)
\textrm{Re}\left(r^{\mu}_{2}\right)}{|1-r^{\mu}_{1}r^{\mu}_{2}\;e^{-y}|^2}.
\label{newhytsdd}
\end{eqnarray}
The relevant frequencies in the integration are $\omega\sim c/l\ll k_BT/\hbar$
 and then it is possible to use the static approximation  for the dielectric functions. In this
approximation the reflection coefficients depends only on one
variable $r^{\mu}_{1,2}(t,y)\rightarrow r^{\mu}_{1,2}(t)$ and one can
reproduce (after the change $y\rightarrow x$) the
asymptotic behavior  (\ref{AsimEW}) for the pressure
$P_{\textrm{th}}^{\textrm{EW}}(T,0,l)$ .

The pressure in the EW sector can be presented in an alternative form
using the multiple-reflection expansion. To this end one can note that
\begin{equation}\label{multex}
    \frac{e^{-xt/\alpha}}{|1-r^{\mu}_{1}r^{\mu}_{2}\;e^{-xt/\alpha}|^2}=
    \sum_{n=1}^{\infty}\frac{\textrm{Im}(r^{\mu}_{1}r^{\mu}_{2})^n}
    {\textrm{Im}(r^{\mu}_{1}r^{\mu}_{2})}\;e^{-nxt/\alpha},
\end{equation}
and can put this expansion in Eq.~(\ref{newhyt}). In the static
approximation the integral over $x$ can be found explicitly:
\begin{equation}\label{xint}
    \int_0^{\infty}\textrm{d}x\frac{x^3}{e^x-1}\;e^{-nxt/\alpha}=\Psi^{(3)}\left(1+nt/\alpha\right),
\end{equation}
where $\Psi^{(3)}\left(1+nt/\alpha\right)$ is the polygamma function
\cite{Abram}. Since $\alpha $ is small, one can take only the
asymptotic of this function, which is
$\Psi^{(3)}\left(1+nt/\alpha\right)\rightarrow 2(\alpha/nt)^3$. Then
the EW pressure can be presented as
\begin{eqnarray}
\lefteqn{\!\!\!\!\!\!\!\!\!\!\!\!\!P_{\textrm{th}}^{\textrm{neq,EW}}(T,0,l)=\frac{k_B T}{4\pi^2 l^3
}\sum_{n=1}^{\infty}\frac{1}{n^3}\int_0^{\infty}
\frac{\textrm{d}t}{t}
\times}\notag\\
&&\;\;\;\;\;\;\;\;\;\sum_{\mu=s,p}\frac{\textrm{Im}\left(r^{\mu}_{1}\right)
\textrm{Re}\left(r^{\mu}_{2}\right)}{\textrm{Im}\left(r^{\mu}_{1}r^{\mu}_{2}\right)}
\textrm{Im}\left(r^{\mu}_{1}r^{\mu}_{2}\right)^n. \label{alterEW}
\end{eqnarray}
This representation is helpful for the analysis of
the rarefied body limit that will be presented in the next section. 

It is interesting to derive also the large distance behavior
(\ref{AsimEWEQ}) for the equilibrium case. One half of the
equilibrium pressure, $P^{\textrm{eq,EW}}_{\textrm{th}}(T,l)/2$, is
equal to the symmetric part of Eq.~(\ref{alterEW}) in respect to the
bodies interchange. Therefore, to get
$P^{\textrm{eq,EW}}_{\textrm{th}}(T,l)$ we have to change in
Eq.~(\ref{alterEW}):
\begin{equation}\label{changeEW}
    \frac{\textrm{Im}\left(r^{\mu}_{1}\right)
\textrm{Re}\left(r^{\mu}_{2}\right)}
{\textrm{Im}\left(r^{\mu}_{1}r^{\mu}_{2}\right)}\rightarrow 1.
\end{equation}
In this case the integrand in Eq.~(\ref{alterEW}) becomes an
analytic function of $t$ with the poles at $t=0$ and at infinity.
The integral can be calculated using the quarter-circle contour of
infinite radius closing the positive real axis and negative
imaginary axis. This is because $it\sim q_z$ must have a positive
real part. Finally the integral is reduced to the quarters of the residues
in the poles and gives:
\begin{eqnarray}
    \lefteqn{\!\!\!\!\!\!\!\!\!\!\!P^{\textrm{eq,EW}}_{\textrm{th}}(T,l)=-\frac{k_B T\zeta(3)}{4\pi l^3}+}\notag\\
 &&\;\;\;\;\;\;\;   \frac{k_B T}{8\pi
l^3}\sum_{n=1}^{\infty}\frac{1}{n^3}\left(\frac{\varepsilon_{10}-1}{\varepsilon_{10}+1}
\frac{\varepsilon_{20}-1}{\varepsilon_{20}+1}\right)^n. 
\label{poles}
\end{eqnarray}
The sum in this expression can be written in the equivalent integral
form so that (\ref{poles}) coincides with Eq.~(\ref{AsimEWEQ}). Note
that only $p$-polarization contributes to the pole at infinity. This is because at infinity
$r^s_m\rightarrow 0$ but $r^p_m\rightarrow
(\varepsilon_{m0}-1)/(\varepsilon_{m0}+1)$ stays finite.
%
%%%%%%%%%%%%%%%%%%%%%%%%%%%%%%%%%%%%%%%%%%%%%%%%%%%%%%%%%%%%%%%%%%%%%%%%%%%%%%%%%
\section{\label{sec:LongDnonad}Pressure between a solid and a diluted body}
%%%%%%%%%%%%%%%%%%%%%%%%%%%%%%%%%%%%%%%%%%%%%%%%%%%%%%%%%%%%%%%%%%%%%%%%%%%%%%%%%
A case of particular interest is the interaction between solid 
and diluted bodies. In fact the first measurement of the
non-equilibrium interaction was done between
an ultracold atomic cloud and a dielectric substrate
\cite{Cornell06}. From the theoretical point of view this case is
the most simple for analytical analysis.

Here we investigate  the pressure between a hot
dielectric substrate of temperature $T$ (body 1) and a gas cloud
(body 2) at large distances. When the second body is very dilute we
can consider the limit $(\varepsilon_2-1)\rightarrow 0$. If both
bodies are at the same temperature $T$, the equilibrium pressure can
be found by expanding Eq.~(\ref{AsimTOTEQ}) on small values of $(\varepsilon_2-1)$.
The leading term is
\begin{equation}
P_{\textrm{th}}^{\textrm{eq}}(T,l) = \frac{k_BT}{16\pi
l^3}\frac{\varepsilon_{10}-1}{\varepsilon_{10}+1}\;(\varepsilon_{20}-1).
\label{LifLDbbab}
\end{equation}
This pressure, valid at the condition (\ref{lTTSummary}), is
proportional to $(\varepsilon_2-1)=4\pi n \alpha$, where  $n$ is the
density of the material $2$ and $\alpha$ is the dipole
polarizability of its constituents (for example atoms). We can see
that the pressure is additive since the additivity would in fact
require a linear dependence on the gas density $n$ and hence on
$(\varepsilon _{20}-1)$.

If one performs first the diluteness limit of the exact
surface-surface pressure, and then  takes the large distance limit, one
obtains very interesting asymptotic behaviors for the PW and EW contributions
respectively \cite{AntezzaPhDthesis}
\begin{eqnarray}
\!\!\!P_{\textrm{th}}^{\textrm{eq,PW}}(T,l)&=&- \frac{(k_BT)^2}{24\;l^2\;c\hbar}\frac{\varepsilon_{10}+1}{\sqrt{\varepsilon_{10}-1}}\;(\varepsilon_{20}-1),
\label{ffsDERRRRSummary}\\
\!\!\!P_{\textrm{th}}^{\textrm{eq,EW}}(T,l) &=&
\frac{(k_BT)^2}{24\;l^2\;
c\hbar}\frac{\varepsilon _{10}+1}{\sqrt{\varepsilon _{10}-1}}\;(\varepsilon_{20}-1).
\label{LevDERRRRSummary}
\end{eqnarray}
In deriving these limits we assumed that $k_BT$ is much
smaller than the lowest dielectric resonance of both the body $1$ and
of the atoms of the dilute body $2$. Such asymptotic behaviors for
the PW and EW components depend on the temperature more strongly
than at equilibrium and decay slower at large distances ($\sim
T^2/l^2$). It is also remarkable  that the PW component of the
surface-rarefied body pressure (\ref{ffsDERRRRSummary}) depends on
the dielectric functions and is repulsive, differently from
attractive nature of the PW component of the surface-surface pressure
(\ref{AsimPWEQ}).

The PW and EW terms (\ref{ffsDERRRRSummary}) and
(\ref{LevDERRRRSummary}) exactly cancel each other, and in order to
find the total pressure one should expand the corresponding
expressions to higher order. The final result is given by
Eq.~(\ref{LifLDbbab}). In configurations out of thermal equilibrium
there will be no longer such peculiar cancellations between the PW
and EW terms. In this case the new asymptotic behavior $\sim
T^2/l^2$ will characterize the total pressure at large distances,
while there will be a transition to a $\sim T/l^3$ behavior at
larger distances.

\begin{figure}[ptb]
\begin{center}
\includegraphics[width=0.45\textwidth]{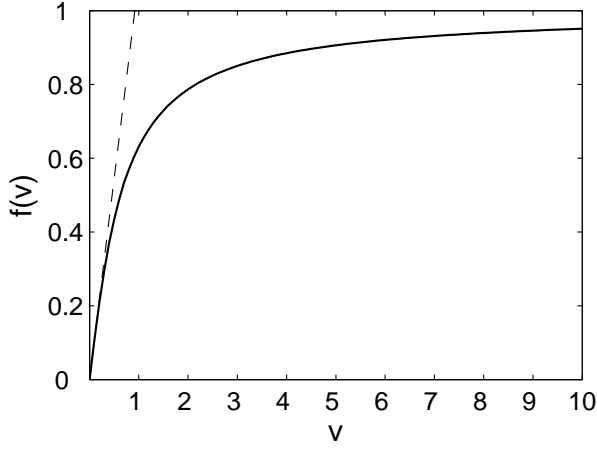}%{fig5.eps}%{transition_function.eps}
\caption{Dimensionless function $f(v)$ [see Eqs.~(\ref{neqrar}) and (\ref{frelat}]
describing the transition between additive and nonadditive regimes.
The dashed line presents the asymptotic limit at small $v$.}
\label{tranfun}
\end{center}
\end{figure}

In particular, the result of the surface-rarefied body pressure out of equilibrium
can be presented as \cite{ShortArticle}
\begin{equation}
    P^{\textrm{neq}}_{\textrm{th}}(T,0,l)=\frac{k_BTC}{l^3}\frac{\varepsilon_{10}+1}
    {\sqrt{\varepsilon_{10}-1}}\sqrt{\varepsilon_{20}-1}f(v),
\label{neqrar}
\end{equation}
where
\begin{equation}\label{xdef}
    v=\frac{l\sqrt{\varepsilon_{20}-1}}{\lambda_T}
\end{equation}
is a dimensionless variable and $C=3.83\cdot10^{-2}$ is a constant. The
function $f(v)$, whose expression will be derived below [see
Eq.(\ref{frelat}), together with  Eqs.~(\ref{fPW}),
(\ref{fEW1}), and (\ref{fEW2})], is a dimensionless function of $v$. It is
possible to show (see derivation below) that $f(v)\rightarrow 1$ for
$v\rightarrow\infty$,  while $f(v)\rightarrow v/24C$  for
$v\rightarrow 0$. This function is shown in Fig.~\ref{tranfun}.
Eq.~(\ref{neqrar}) is valid at the condition of large distances
$l/\lambda_T\gg 1$, which does not restrict the value of $v$.
%
%===================================================================================================
%===================================================================================================

At large values of $v$ the pressure (\ref{neqrar}) becomes
\begin{equation}
    P^{\textrm{neq}}_{\textrm{th}}(T,0,l)=\frac{k_BTC}{l^3}\frac{\varepsilon_{10}+1}
    {\sqrt{\varepsilon_{10}-1}}\sqrt{\varepsilon_{20}-1},
\label{neqrardde}
\end{equation}
and is proportional to $\sqrt{\varepsilon _{20}-1}$. This peculiar
dependence means that the pressure acting on the atoms of the
substrate $2$ \textit{is not additive}. The non additivity of the
pressure can be physically explained as follow:  for large $l$ the main
contribution to the force is produced by the \textit{grazing waves}
incident on the interface of the material $2$ from the vacuum gap
with small values of $q_z/k\leq \sqrt{\varepsilon _{20}-1}$. Hence the reflection
coefficients from the body $2$ is not small even at small
$\varepsilon _{20}-1$ and the body cannot be considered as dilute
from an electrodynamic point of view \cite{ShortArticle}. This is a peculiarity of the
non-equilibrium situation. In fact at equilibrium this anomalous
contribution is canceled by the waves impinging the interface  from
the interior of the dielectric $2$, close to the angle of total
reflection. In a rarefied body such waves become grazing. Notice
that the pressure (\ref{neqrardde}) is valid at the condition
\begin{equation}
l\gg \frac{\lambda_T}{\sqrt{\varepsilon_{20}-1}},
\label{largel}
\end{equation}
which  becomes stronger and stronger as $(\varepsilon _{20}-1)~\to~0$.

At small $v$ one finds from Eq.(\ref{neqrar})
\begin{equation}
    P^{\textrm{neq}}_{\textrm{th}}(T,0,l)=\frac{(k_BT)^2}{24 l^2\hbar c}\frac{\varepsilon_{10}+1}
    {\sqrt{\varepsilon_{10}-1}}(\varepsilon_{20}-1).
\label{smallx}
\end{equation}
In this case the additivity is restored but the temperature
dependence is not linear any more and the pressure decreases more
slowly with the distance. This result holds at distances
\begin{equation}
    \lambda_T\ll l\ll \frac{\lambda_T}{\sqrt{\varepsilon_{20}-1}}.
\label{smallxrange}
\end{equation}
It is worth noting that the interval (\ref{smallxrange}) practically
disappears for dense dielectrics. 

\begin{figure}[ptb]
\begin{center}
\includegraphics[width=0.45\textwidth]{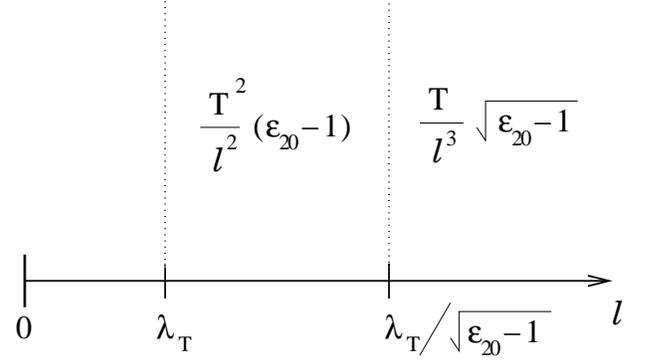}
\caption{Relevant length scales and asymptotic behaviors of the surface-rarefied body pressure out of thermal equilibrium. There is a first region given by Eq.(\ref{smallxrange}) where the pressure is additive and coincides with Eq.(\ref{smallx}), and a second region, satisfying Eq.(\ref{largel}), where the pressure is given by Eq.(\ref{neqrardde}) and is no longer additive.}
\label{FigCross}
\end{center}
\end{figure}

The above discussion  can be summarized as follow [see Fig.\ref{FigCross}]. If the dielectric $2$ is very dilute but still occupies an infinite half space (or anyway is thick enough, in the sense defined above), there is a first region given by Eq.(\ref{smallxrange}) where the pressure is additive and coincides with Eq.(\ref{smallx}). At larger distances, satisfying Eq.(\ref{largel}), the pressure is given by Eq.(\ref{neqrardde}) and is no longer additive. In the intermediate region $l\sim \lambda_T/\sqrt{\varepsilon _{20}-1} $ equations (\ref{smallx}) and (\ref{neqrardde}) are of the same order. 
%===================================================================================================
%===================================================================================================
\\

It is interesting to note that, due to the diluteness condition
$(\varepsilon_{20}-1)\ll1$, in both regions (\ref{largel})
and (\ref{smallxrange}) the thermal term $\Delta P_{\textrm{th}}$ [sum of Eqs.(\ref{EWI}) and (\ref{PWA22})]
gives the leading contribution into the $l-$dependent component of
the total pressure $P^{\textrm{neq}}_{\textrm{th}}(T,0,l)$. 
This clearly emerges from Eqs.(\ref{neqPW2}) - (\ref{neqEW2}), by comparing (fot $T_2=0$) the large distance behavior of the pressure at equilibrium $P^{\textrm{eq}}_{\textrm{th}}(T,l)$ given by Eq.(\ref{LifLDbbab}), with the large distance behaviors of the total pressure just derived, given by Eqs.(\ref{neqrardde}) and (\ref{smallx}). The consequences of this are remarkable. In fact the large distance behavior of the total pressure becomes proportional to   $$P^{\textrm{neq}}_{\textrm{th}}(T_1,T_2,l)\simeq\Delta P_{\textrm{th}}(T_1,l)-\Delta P_{\textrm{th}}(T_2,l),$$ and  {\it the interaction between the two bodies
will be attractive if} $T_1>T_2$ {\it and repulsive in the opposite case}
\cite{ShortArticle}.\\

Below, in sections \ref{pwrarsec} and \ref{EWrarsec}, we present the derivation of
 Eq.(\ref{neqrar}) for both the PW and EW
components, which give rise respectively to the asymptotic behaviors
(\ref{neqrardde}) and (\ref{smallx}).
%
%%%%%%%%%%%%%%%%%%%%%%%%%%%%%%%%%%%%%%%%%%%%%%%%%%%%%%%%%%%%%%%%%%%%%%%%%%%%%%%%%%%%%%%%%%%
\subsection{\label{pwrarsec}PW contribution}
%%%%%%%%%%%%%%%%%%%%%%%%%%%%%%%%%%%%%%%%%%%%%%%%%%%%%%%%%%%%%%%%%%%%%%%%%%%%%%%%%
In this section we focus on  the PW contribution to
Eq.~(\ref{neqrar}). One can do explicit calculations if the
dielectric functions of the materials do not depend on frequency.
This is a good approximation for the diluted body 2. Since we are
interested in the large distance asymptotic, this approximation is
also good for the solid body 1 if the material has no resonances for
$\omega<c/l\ll k_BT/\hbar$. In the case of static dielectric
functions the integral over $x$ in Eq.~(\ref{ppasde}) can be
evaluated via the polygamma function:
\begin{equation}\label{xintimag}
    \int_0^{\infty}\frac{\textrm{d}xx^3}{e^x-1}e^{intx/\alpha}=\Psi^{(3)}
    \left(1-i\frac{nt}{\alpha}\right).
\end{equation}
Then, introducing the new variable $u$ instead of $t$ and the
parameter $b$ according to the definitions
\begin{equation}\label{ubdef}
    u=\frac{t}{\sqrt{\varepsilon_{20}-1}},\ \ \
    b=\sqrt{\frac{\varepsilon_{10}-1}{\varepsilon_{20}-1}}\gg1
\end{equation}
one can expand $r_1^{\mu}(u)$ in series of $1/b$
\begin{equation}\label{r1exp}
    r_1^s\approx -\left(1-\frac{2u}{b}\right),\ \ \
    r_1^p\approx -\left(1-\frac{2\varepsilon_1 u}{b}\right)
\end{equation}
and in the same approximation one has
\begin{equation}\label{r2u}
    r_2^p\approx r_2^s=r_2=\frac{u-\sqrt{1+u^2}}{u+\sqrt{1+u^2}}.
\end{equation}
The result is the following expression for the
pressure $P_{\textrm{th}}^{\textrm{neq,PW}}(T,0,l)$:
\begin{eqnarray}
   P_{\textrm{th}}^{\textrm{neq,PW}}(T,0,l)=
    -\frac{2(k_B T)^4}{\pi^2\hbar^3 c^3}
    \frac{\varepsilon_{10}+1}{\sqrt{\varepsilon_{10}-1}}(\varepsilon_{20}-1)^{2}\sum_{n=1}^{\infty}\times\notag\\
\int_{0}^{1/\sqrt{\varepsilon_{20}-1}}\textrm{d}uu^3
    \frac{1+r_2^2}{1-r_2^2}(-r_2)^n\textrm{Re}\Psi^{(3)}\left(1-i2nvu\right),\notag\\
    \label{neqPW7A}
\end{eqnarray}
where the parameter $v$ is given by Eq.~(\ref{xdef}). Here the sum
on polarizations gave the factor $\varepsilon_{10}+1$. In the
leading approximation the integration over $u$ can be extended up to
infinity. Furthermore, the real part of
$\Psi^{(3)}\left(1-iy\right)$ can be presented as \cite{Abram}
\begin{equation}\label{real3}
    \textrm{Re}\Psi^{(3)}\left(1-iy\right)=\frac{\pi}{2}\frac{\textrm{d}^3}
    {\textrm{d}y^3}\left(\frac{1}{\pi y}-\coth\pi
    y\right).
\end{equation}
After some transformations equation (\ref{neqPW7A}) becomes:
\begin{equation}\label{PWcontr}
    P_{\textrm{th}}^{\textrm{neq,PW}}(T,0,l)=\frac{k_B T}{l^3}
    \frac{\varepsilon_{10}+1}{\sqrt{\varepsilon_{10}-1}}\sqrt{\varepsilon_{20}-1}f_{PW}(v),
\end{equation}
where $f_{PW}(v)$ is given by

\begin{eqnarray}
    \lefteqn{\!\!\!\!\!\!\!\!\!\!\!\!\!\!\!\!\!\!\!\!\!\!\!\!f_{PW}(v)=-\frac{1}{8\pi}\sum_{n=1}^{\infty}\frac{1}{n^3}
    \int_{0}^{\infty}\textrm{d}uu^3
    \frac{1+r_2^2}{1-r_2^2}(-r_2)^n\times}\notag\\
    &&\!\!\!\frac{\textrm{d}^3}
    {\textrm{d}u^3}\left[\frac{1}{2\pi nvu}-\coth(2\pi nv
    u)\right].
    \label{fPW}
\end{eqnarray}
The function $f_{PW}(v)$ can be calculated explicitly for large
and small values of  $v$. When $v\gg 1$, the important range of $u$
in the integral (\ref{fPW}) is $u\ll 1$ and one can expand the
reflection coefficient $r_2$ on small values of $u$. Then the
function $f_{PW}(v)$ is reduced to
\begin{eqnarray} \lefteqn{\!\!\!\!\!\!\!\!\!\!\!\!\!\!\!\!\!\!\!\!\!\!\!\!f_{PW}(v\rightarrow\infty)=-\frac{1}{16\pi}\sum_{n=1}^{\infty}\frac{1}{n^3}
    \int_{0}^{\infty}\textrm{d}uu^2
    \times}\notag\\
    &&\!\!\!\frac{\textrm{d}^3}
    {\textrm{d}u^3}\left[\frac{1}{2\pi nvu}-\coth(2\pi nv
    u)\right].\label{fPWlarge}
\end{eqnarray}
The integral here is easily calculated by parts and finally one finds
\begin{equation}\label{fPWLarge}
    f_{PW}(v\rightarrow\infty)=C_{PW}=\frac{\zeta(3)}{8\pi}.
\end{equation}

\begin{figure}[ptb]
\begin{center}
\includegraphics[width=0.45\textwidth]{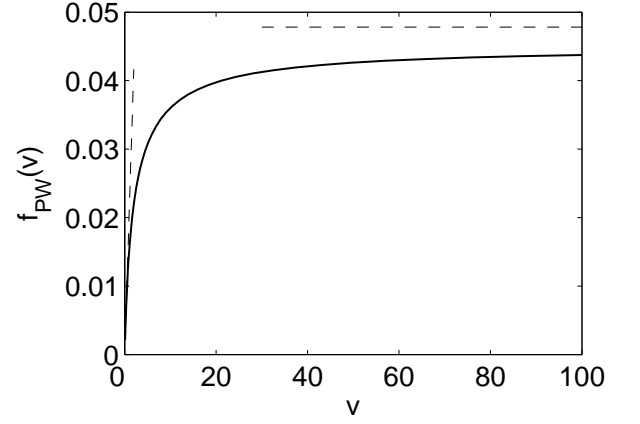}%{fig6.eps}
\caption{Function $f_{PW}(v)$ (solid) [Eq.(\ref{fPW})] and its asymptotic limits (dashed) at small [Eq.(\ref{fPWSmall})]
and large [Eq.(\ref{fPWLarge})] values of $v$.} \label{fig6}
\end{center}
\end{figure}

When $v\ll 1$ the significant values of $u$ in the integral
(\ref{fPW}) are $u\gg 1$, and one can make the corresponding
expansion in the reflection coefficient (\ref{r2u}). In this case
only the $n=1$ term in the sum is relevant. Then one obtains
\begin{eqnarray}\label{fPWsmall}
    \lefteqn{\!\!\!\!\!\!\!\!\!\!\!\!\!\!\!\!\!\!f_{PW}(v\rightarrow 0)=-\frac{1}{32\pi}\int_{0}^{\infty}\textrm{d}uu\times}\notag\\
    &&\!\!\!\!\!\!\frac{\textrm{d}^3}
    {\textrm{d}u^3}\left[\frac{1}{2\pi vu}-\coth(2\pi v
    u)\right],
\end{eqnarray}
and finally:
\begin{equation}\label{fPWSmall}
    f_{PW}(v\rightarrow 0)=\frac{v}{48}.
\end{equation}
The function $f_{PW}(v)$ and its asymptotic behaviours at large and small $v$
are shown in Fig.\ref{fig6}.

%===================================================================================================
%===================================================================================================

%%%%%%%%%%%%%%%%%%%%%%%%%%%%%%%%%%%%%%%%%%%%%%%%%%%%%%%%%%%%%%%%%%%%%%%%%%%%%%%%%%%%%%%%%%%
\subsection{\label{EWrarsec}EW contribution}
%%%%%%%%%%%%%%%%%%%%%%%%%%%%%%%%%%%%%%%%%%%%%%%%%%%%%%%%%%%%%%%%%%%%%%%%%%%%%%%%%
The derivation of the EW component of the pressure (\ref{neqrar})
can be performed starting from the expression (\ref{newhyt}) for the
pressure $P_{\textrm{th}}^{\textrm{neq,EW}}(T,0,l)$. By performing
the multiple-reflection expansion with the help of
Eq.~(\ref{multex}) and calculating the integral over the variable
$x$ using (\ref{xint}) one finds that Eq.(\ref{newhyt}) becomes
\begin{eqnarray}\label{neqEWrar}
    \lefteqn{\!\!\!\!\!\!\!P_{\textrm{th}}^{\textrm{neq,EW}}(T,0,l)=
    \frac{(k_B T)^4}{\pi^2\hbar^3 c^3}
    \sum_{n=1}^{\infty}\int_{0}^{\infty}\textrm{d}tt^2\times}\notag\\
    &&\sum_{\mu=s,p}
    \frac{\textrm{Im}r_1^{\mu}\textrm{Re}r_2^{\mu}}{\textrm{Im}(r_1^{\mu}r_2^{\mu})}
    \textrm{Im}(r_1^{\mu}r_2^{\mu})^n\Psi^{(3)}\left(1+\frac{nt}{\alpha}\right),
\end{eqnarray}
As in the case of the propagating waves one can introduce the variable
$u$ instead of $t$ according to Eq.~(\ref{ubdef}), and can make the expansion
for large $b$. Then for the reflection coefficients one gets
\begin{eqnarray}\label{reflEW}
  r_1^s\approx -\left(1-\frac{i2u}{b}\right),\ \ \
    r_1^p\approx -\left(1-\frac{i2\varepsilon_1 u}{b}\right)\notag\\
  r_2^p\approx r_2^s = \frac{iu-\sqrt{1-u^2}}{iu+\sqrt{1-u^2}}.
  \ \ \ \ \ \ \ \ \ \
\end{eqnarray}
Now one should distinct the integration ranges $0<u<1$ and
$1<u<\infty$, since the integrands are different in these ranges.
Let us do it with the superscript $(1)$ or $(2)$, respectively.

As in the case of propagating waves (\ref{PWcontr}) the pressure can
be presented as a parameter-dependent factor times a universal
function of $v=l\sqrt{\varepsilon_{20}-1}/\lambda_T$:
\begin{equation}\label{EWcontr}
    P_{\textrm{th}}^{\textrm{neq,EW}}(T,0,l)=\frac{k_B T}{l^3}
    \frac{\varepsilon_{10}+1}{\sqrt{\varepsilon_{10}-1}}\sqrt{\varepsilon_{20}-1}f_{EW}(v),
\end{equation}
where the function $f_{EW}(v)$ includes contributions from $0<u<1$
and $1<u<\infty$ ranges:
\begin{equation}\label{2contr}
    f_{EW}(v)=f_{EW}^{(1)}(v)+f_{EW}^{(2)}(v).
\end{equation}
For these functions one has the following
expressions
\begin{eqnarray}\label{fEW1}
    \!\!\!\!\!\!\!\!\!\!f_{EW}^{(1)}(v)=-\frac{1}{4\pi^2}\sum_{n=1}^{\infty}\frac{1}{n^3}
    \int_{0}^{1}\textrm{d}uu^3
    \times\notag\\
    \frac{2u^2-1}{2u\sqrt{1-u^2}}\textrm{Im}(-r_2)^n\frac{\textrm{d}^3}
    {\textrm{d}u^3}\Psi\left(1+2nvu\right);
\end{eqnarray}
\begin{eqnarray}\label{fEW2}
    f_{EW}^{(2)}(v)=-\frac{1}{4\pi^2}\sum_{n=1}^{\infty}\frac{1}{n^3}
    \int_{1}^{\infty}\textrm{d}uu^3(-r_2)^n\times\notag\\
    \frac{\textrm{d}^3}
    {\textrm{d}u^3}\Psi\left(1+2nvu\right).
\end{eqnarray}
Here we have used the relation between the polygamma functions
\begin{equation}\label{pgrel}
    \Psi^{(3)}(1+y)=\frac{\textrm{d}^3}{\textrm{d}y^3}\Psi\left(1+y\right),
\end{equation}
where $\Psi\left(1+y\right)$ is the digamma function \cite{Abram}.

Let us discuss now the asymptotic behavior of the functions
$f_{EW}^{(1),(2)}(v)$ at small and large values of $v$. For large
$v$ the contribution from the range $u\lesssim 1/v$ in the integral
Eq.~(\ref{fEW1}) is negligible,  and one can consider the digamma
function at large arguments $\Psi(1+2nvu)\rightarrow \ln(2nvu)$.
Then the integral can be calculated after the substitution
$u=\sin\varphi$. It gives the following result
\begin{eqnarray}\label{fEW1Large} 
\lefteqn{\!\!\!\!\!\!\!f_{EW}^{(1)}(v\rightarrow\infty)=
-\frac{1}{4\pi^2}\sum_{n=1}^{\infty}\frac{1}{n^3}\times}\notag\\
    &&\left[\frac{\pi}{2}+(-1)^n\left(\beta(n+3/2)+\frac{2}{4n^2-1}\right)\right],
\end{eqnarray}
where the $\beta$-function is defined as
\begin{equation}\label{beta}
    \beta(y)=\frac{1}{2}\left[\Psi\left(\frac{1+y}{2}\right)-\Psi\left(\frac{y}{2}\right)\right].
\end{equation}
To find $f_{EW}^{(2)}(v)$ one can also take the asymptotic value of
$\Psi(1+2nvu)$, make the change $u=\cosh\chi$, and after the
integration one obtains
\begin{eqnarray}\label{fEW2Large}
    f_{EW}^{(2)}(v\rightarrow\infty)=-\frac{1}{4\pi^2}\sum_{n=1}^{\infty}
    \frac{2(-1)^n}{n^3(4n^2-1)}.
\end{eqnarray}
Taking the sum of both functions $f_{EW}^{(1)}$ and $f_{EW}^{(2)}$
one finds finally the large $v$ asymptotic for $f_{EW}(v)$:
\begin{eqnarray}\label{fEWLarge}
    \lefteqn{\!\!\!\!\!\!\!f_{EW}(v\rightarrow\infty)=-\frac{1}{4\pi^2}\sum_{n=1}^{\infty}\frac{1}{n^3}\times}\notag\\
    &&\left[\frac{\pi}{2}+(-1)^n\left(\beta(n+3/2)+\frac{2(n+1)}{4n^2-1}\right)\right].
\end{eqnarray}
This sum is just a number equal to
\begin{equation}\label{CEW}
    f_{EW}(v\rightarrow\infty)=C_{EW}=-0.96\cdot10^{-2}.
\end{equation}
Combining together the large $v$ contributions from PW
(\ref{fPWLarge}) and EW (\ref{CEW}) one can find the constant in
Eq.~(\ref{neqrar}), i.e. $C=C_{PW}+C_{EW}=3.83\cdot10^{-2}$.

In the limit of small $v$ it is not difficult to show that
$f_{EW}^{(1)}(v)\sim v^3$ and can be neglected. The main
contribution to $f_{EW}^{(2)}(v)$ comes from the range $u\sim 1/v\gg
1$. For these values the reflection coefficient $r_2(u)\approx
1/4u^2$ is small and only the $n=1$ term in the sum is relevant.
Then the integral over $u$ can be calculated by parts and one
obtains
\begin{equation}\label{fEWSmall}
    f_{EW}(v\rightarrow 0)=\frac{v}{48}.
\end{equation}
Let us note that for small $v$, the PW and EW contributions
coincide.

\begin{figure}[ptb]
\begin{center}
\includegraphics[width=0.45\textwidth]{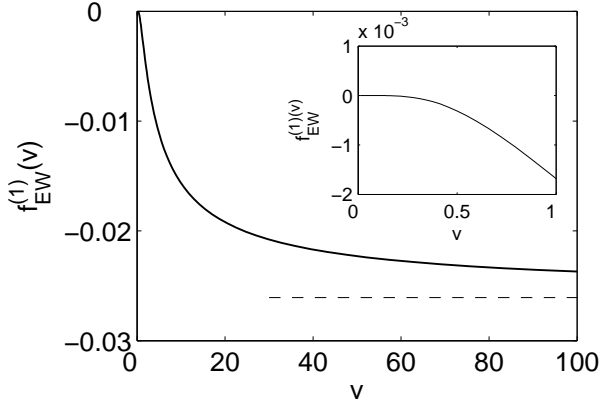}%{fig7.eps}
\caption{Function $f^{(1)}_{EW}(v)$ (solid) [Eq.(\ref{fEW1})]. The asymptotic limit at large values of $v$
[Eq.(\ref{fEW1Large})] is shown by the dashed line. The inset demonstrates $v^3$
behavior at small $v$.} \label{fig7}
\end{center}
\end{figure}

\begin{figure}[ptb]
\begin{center}
\includegraphics[width=0.45\textwidth]{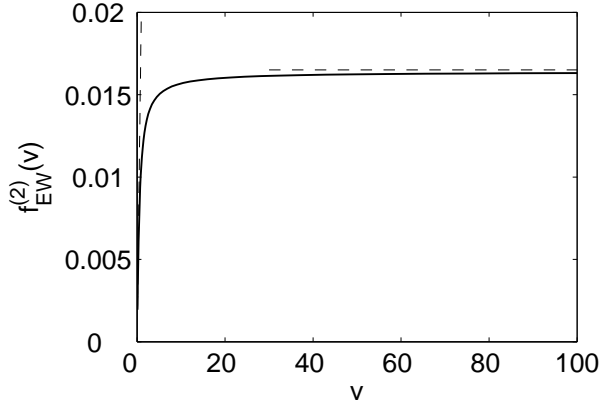}%{fig8.eps}
\caption{Function $f^{(2)}_{EW}(v)$ (solid) [Eq.(\ref{fEW2})] and its asymptotic limits (dashed) at
small and large [Eq.(\ref{fEW2Large})] values of $v$.} \label{fig8}
\end{center}
\end{figure}

The function $f^{(1)}_{EW}(v)$ is shown in Fig.\ref{fig7}. The inset
demonstrates the cubic behavior at small $v$. It should be noted that
$f^{(1)}_{EW}(v)$ so as $f_{PW}(v)$ approach the large $v$
asymptotics rather slowly, but the sum of these functions reaches
the large $v$  limit faster as Fig.\ref{tranfun} demonstrates. The
function $f^{(2)}_{EW}(v)$ is presented in Fig.\ref{fig8}. One can
see that it behaves in accordance with expected asymptotics.

%===================================================================================================
%===================================================================================================

%===================================================================================================
%===================================================================================================

Finally, one can establish the correspondence between the function
$f(v)$ entering the general formula (\ref{neqrar}) for the pressure
in the limit of one diluted body and the functions $f_{PW}(v)$,
$f_{EW}^{(1)}(v)$, and $f_{EW}^{(2)}(v)$ given by Eqs.~(\ref{fPW}),
(\ref{fEW1}), and (\ref{fEW2}), respectively. This correspondence is
given by the simple relation
\begin{equation}
    C\;f(v)=f_{PW}(v)+f_{EW}^{(1)}(v)+f_{EW}^{(2)}(v).
\label{frelat}
\end{equation}
%
%
%%%%%%%%%%%%%%%%%%%%%%%%%%%%%%%%%%%%%%%%%%%%%%%%%%%%%%%%%%%%%%%%%%%%%%%%%%%%%%%%%%%%%%%%%%%
\section{\label{AtSurfNeq}Large distance behavior of the surface-atom force out of thermal equilibrium}
%%%%%%%%%%%%%%%%%%%%%%%%%%%%%%%%%%%%%%%%%%%%%%%%%%%%%%%%%%%%%%%%%%%%%%%%%%%%%%%%%
It is interesting to recover the asymptotic results of the
surface-atom force out of thermal equilibrium (obtained in
\cite{articolo2}) from the general expression of the pressure
given by Eqs.(\ref{neqPW}) and (\ref{neqEW}). To do this it is crucial to
carry out the limit $(\varepsilon_{2}-1)=4\pi n\alpha_2 \rightarrow
0$ before taking the limit of large distances. To show this, let us
focus first on the EW term given by Eq.(\ref{neqEW}), and perform
the rarefied body expansion  (body 2) assuming that
$\sqrt{\varepsilon_{20}-1}$ is the smallest quantity, also with
respect to $|q_z|/k$. Due to the effect of the Bose factor, only the
frequencies $\omega\sim k_B T/\hbar$ are relevant in the
integration, and due to the exponential $e^{-2l|q_z|}$ the relevant
wave-vectors are given by
\begin{equation}
|q_z|/k\sim \lambda_T/l\gg\sqrt{\varepsilon_{20}-1}.
\label{additivity}
\end{equation}
In this way at large distance  it is easy to reproduce the Eqs.(10)-(11) of \cite{articolo2}:
\begin{eqnarray}
\lefteqn{\!\!\!\!\!\!\!\!\!\!\!{P}_{\text{th}}^{\text{neq,EW}}(T,0,l)=\frac{\hbar \left( \varepsilon
_{20} -1\right) }{l^{2}\;8\pi ^{2}\;c}\int_{0}^{\infty }\text{d}%
\omega \frac{\omega }{e^{\hbar \omega /k_BT}-1}\;\times}\notag\\
&&\!\!\!\sqrt{ |\varepsilon_{1}(\omega )-1|+\left[\varepsilon
_{1}^{\prime }(\omega )-1\right]}\;\frac{2+|\varepsilon
_{1}(\omega )-1|}{\sqrt{2}|\varepsilon _{1}(\omega )-1|}.
  \label{f}
\end{eqnarray}
In deriving Eq.(\ref{f}) we also replaced $\varepsilon _{2}\left(
\omega \right) $ with its static value $\varepsilon _{20}$, which is
reasonable if $k_BT$ is much smaller than the lowest atomic
resonances, and also  ensures that the atoms of the dilute body $2$
cannot adsorb the thermal radiation.

 For a rarefied body one has that $\varepsilon _{20}-1\approx 4\pi \alpha _{0}n_{a}$,
where $n_{a}$ is the number of atoms of the body per unit volume and $%
\alpha _{0}$ is the static polarizability of an atom. The pressure in this
case is proportional to $n_{a}$ and the force acting on an individual atom
can be calculated as
\begin{equation}
F_{{\rm th}}^{{\rm neq,EW}}=\frac{1}{n_{a}}\frac{dP_{{\rm th}}^{{\rm neq,EW}}%
}{dl}.  \label{peratom}
\end{equation}
It is easy to check that substituting Eq.(\ref{f}) into (\ref{peratom}) one obtains
exactly Eq.(10) of \cite{articolo2}.

However, there is also the PW contribution. The expansion in the $l$
-dependent part of the PW pressure (\ref{neqPW}) produces a
contribution identical to the EW one, thereby doubling the value of
the force (\ref{peratom}). This apparent contradiction can be easily
solved by the following arguments. The problem approached in the
present paper is not equivalent from that approached in
Ref.~\cite{articolo2}. Here we assume that the second slab, being
rarefied, is still thick enough to absorb black body radiation from the
first slab. On the contrary, the transition to individual atoms
(which is the case discussed in Ref.~\cite{articolo2}) demands
to completely neglect the absorption. Then, to calculate the surface-atom force correctly, one must
consider the limit $\varepsilon _{2}^{\prime \prime }\rightarrow 0$
at finite thickness $L$ of the slab 2. On the contrary using the expression
(\ref{neqPW}) means taking the opposite limit procedure, i.e. first
$L\rightarrow \infty $ and later $\varepsilon _{2}^{\prime \prime
}\rightarrow 0$.  The reason why the first limiting procedure is correct in this case, is that if the slab $2$
does not absorb radiation completely, one should also take into
account the pressure acting on the remote surface (i.e. the external one), generated by the radiation coming from the left. In absence of
absorption it is possible to show that the inclusion of the remote
surface in the slab $2$ results in a relatively small value of the PW pressure.
%, of the order $\left( \sqrt{%
%\varepsilon _{20}-1}l/\lambda _{T}\right) ^{4}$ with respect to Eq.(\ref{f}).
Details of calculations are presented in the Appendix \ref{sec:GLevs}. We only notice here
that neglecting of absorption actually requires the condition $\varepsilon
_{2}^{\prime \prime }\ll \lambda _{T}^{2}/lL$.

As a consequence, for a finite slab of rarefied gas without
absorption the EW contribution (\ref{f})  provides the total
pressure and is equivalent to  equations (10)-(11) of
\cite{articolo2} for the surface-atom force. In particular at
temperatures less than the lowest resonance in
$\varepsilon_1(\omega)$ the pressure (\ref{f}) (and hence the total
pressure) takes the form \cite{articolo2}
\begin{equation}
P_{\text{th}}^{\text{neq,EW}}(T,0,l)= \frac{(k_BT)^2}{48\;l^{2}\;
c\hbar}\frac{\varepsilon _{10}+1}{\sqrt{\varepsilon _{10}-1}}\; (\varepsilon_{20}-1).
\label{art2}
\end{equation}
The above result holds at distances (\ref{smallxrange}).
%
%%%%%%%%%%%%%%%%%%%%%%%%%%%%%%%%%%%%%%%%%%%%%%%%%%%%%%%%%%%%%%%%%%%%%%%%%%%%%%%%%
\section{\label{conclusions}Conclusions}
%%%%%%%%%%%%%%%%%%%%%%%%%%%%%%%%%%%%%%%%%%%%%%%%%%%%%%%%%%%%%%%%%%%%%%%%%%%%%%%%%
In this paper we generalized the Casimir-Lifshitz theory for the
surface-surface pressure to a situation out of thermal equilibrium, when two
bodies are kept at different temperatures in a stationary configuration. In contrast with the equilibrium case, the
non-equilibrium force cannot be presented as the sum over imaginary
frequencies and one has to work in the real frequency domain. At
real frequencies it is natural to separate contributions from
propagating and evanescent waves. The delicate interplay between these
contributions set the total force.

For bodies made of similar materials the pressure is expressed via
the forces at equilibrium. In the general case there is an
additional contribution to the pressure, which is antisymmetric in
respect to interchange of the materials. The propagating part of the
force contains distance independent terms, due to the presence of an energy flux between the bodies in absence of
equilibrium.

We presented a detailed analysis of the force, with particular attention  for large separations/high temperature behaviors. At equilibrium significant cancellations between
PW and EW contributions occur. Such cancellations are less
pronounced in the non-equilibrium situation. It is  established that
at large distances the force between heated ($T$) and cold ($T=0$)
bodies behaves similar to the Lifshitz limit, $\sim T/l^3$, but with
different numerical coefficient. However, this result is true
only for dense bodies. If one of them is diluted the behavior
of the force can change.

Special attention was devoted to the case when one body is diluted.
This is an important situation from which one can recover the interaction
between a body and a single atom. Two remarkable results are found for this
situation \cite{ShortArticle}. First, at very large distances,
$l\gg\lambda_T/\sqrt{\varepsilon_{20}-1}$,
the pressure becomes non-additive, in contrast with the equilibrium
case. Namely, the non-equilibrium pressure is proportional to the
square root of the density of the diluted body, while in the
equilibrium it is proportional to the first power of the density
and, therefore, it is additive. The second result concerns
smaller distances, $\lambda_T\ll
l\ll\lambda_T/\sqrt{\varepsilon_{20}-1}$. In this case we found
a new asymptotic behavior for the pressure, $\sim T^2/l^2$, that decays
with the distance more slowly than the Lifshitz limit at equilibrium, and has a stronger temperature dependence. A careful
analysis of the transition region between these two limits was
done both analytically and numerically.

The pressure between diluted and dense bodies in the distance range
$\lambda_T\ll l\ll\lambda_T/\sqrt{\varepsilon_{20}-1}$ is used to
deduce the surface-atom force. Earlier and with different methods it
was found in \cite{articolo2} that at large distances this force
must behave as $\sim T^2/l^2$. The direct transition from the case
of the surface-diluted body provides a force which is two times larger than
that in Ref.~\cite{articolo2}, and both EW and PW terms contribute
in the same way. We provided a detailed explanation why if the atom does not absorb radiation one has to neglect the contribution of the PW
term, hence recovering the known result.
%
%
%%%%%%%%%%%%%%%%%%%%%%%%%%%%%%%%%%%%%%%%%%%%%%%%%%%%%%%%%%%%%%%%%%%%%%%%%%%%%%%%%
\section{\label{Ack}Acknowledgments}
%%%%%%%%%%%%%%%%%%%%%%%%%%%%%%%%%%%%%%%%%%%%%%%%%%%%%%%%%%%%%%%%%%%%%%%%%%%%%%%%%
We acknowledge supports by the INFN-MICRA project and the Ministero dell'Istruzione, dell' Universit\'a e della Ricerca (MiUR).
\appendix
%%%%%%%%%%%%%%%%%%%%%%%%%%%%%%%%%%%%%%%%%%%%%%%%%%%%%%%%%%%%%%%%%%%%%%%%%%%%%%%%%
\section{\label{sec:GTomas}Green functions for two parallel dielectric half-spaces}
%%%%%%%%%%%%%%%%%%%%%%%%%%%%%%%%%%%%%%%%%%%%%%%%%%%%%%%%%%%%%%%%%%%%%%%%%%%%%%%%%
In this section we present the Green function, which is a solution
of Eq.(\ref{elweqcfe}). We use the Sipe Green-function formalism
\cite{Sipe} for surface optics. Sipe formulated the problem in terms
of $s-$ and $p-$polarized EM vectors waves, in of the
Fresnel coefficients of the interfaces. Here
we use the lateral Fourier transform representation for the Green's
function:
\begin{gather}
G_{ij}\left[\omega;{\bf r},{\bf
r}'\right]=\int\frac{\textrm{d}^2{\bf Q}}{(2\pi)^2} \;e^{i{\bf
Q}\cdot({\bf R}-{\bf R}')}\;g_{ij}\left[\omega;{\bf Q},z,z'\right].
\label{;lkjfpwei}
\end{gather}
In our geometry the Fourier transform $g_{ij}\left[\omega;{\bf
Q},z,z'\right]$ depends only from the modulus $Q=|{\bf Q}|$.
%
%%%%%%%%%%%%%%%%%%%%%%%%%%%%%%%%%%%%%%%%%%%%%%%%%%%%%%%%%%%%%%%%%%%%%%%%%%%%%
\subsection{\label{sec:MMEss}Green's function with the source and the observation points in the vacuum gap}
%%%%%%%%%%%%%%%%%%%%%%%%%%%%%%%%%%%%%%%%%%%%%%%%%%%%%%%%%%%%%%%%%%%%%%%%%%%%%%%%%
If both the observation point ${\bf r}$ and the source point ${\bf
r}'$ are in the vacuum gap, the Green function can be written as the
sum $G_{ij}\left[\omega;{\bf r},{\bf
r}'\right]=G_{ij}^{\textrm{sc.}}\left[\omega;{\bf r},{\bf
r}'\right]+G_{ij}^{\textrm{bu.}}\left[\omega;{\bf r},{\bf
r}'\right]$, of a \textit{scattered} and \textit{bulk} part. In
particular the Fourier transform of these terms are \cite{Tomas95}:
\begin{widetext}
\begin{eqnarray}
g_{ij}^{\textrm{sc.}}\left[\omega;{\bf Q},z,z'\right]&=&\frac{2\pi i k^2}{q_z}\sum_{\mu=s,p}\frac{1}{D_{\mu}}\left[{ e}_{\mu,i}(+)\;{ e}_{\mu,j}(+)\;r_{1}^{\mu}r_{2}^{\mu}\;e^{iq_z(z-z'+2l)}+{ e}_{\mu,i}(+)\;{ e}_{\mu,j}(-)\;r_{1}^{\mu}\;e^{iq_z(z+z')}+\right.\notag\\
&&\left.{ e}_{\mu,i}(-)\;{ e}_{\mu,j}(+)\;r_{2}^{\mu}\;e^{-iq_z(z+z'-2l)}+\right.
\left.{ e}_{\mu,i}(-)\;{ e}_{\mu,j}(-)\;r_{1}^{\mu}r_{2}^{\mu}\;e^{-iq_z(z-z'-2l)}\right],
\label{pppeeepep}\\
\notag\\
g_{ij}^{\textrm{bu.}}\left[\omega;{\bf Q},z,z'\right]&=&-4\pi\delta_{i3}\delta_{j3}\delta(z-z')+\notag\\
&&\frac{2\pi i k^2}{q_z}\sum_{\mu=s,p}\left[{e}_{\mu,i}(+)\;{ e}_{\mu,j}(+)\;e^{iq_z(z-z')}\;\theta(z-z')+{e}_{\mu,i}(-)\;{ e}_{\mu,j}(-)\;e^{-iq_z(z-z')}\;\theta(z'-z)\right].
\label{pppebulk}
\end{eqnarray}
\end{widetext}
Here the multiple reflections enter only in the scattered term and are described by the denominator
\begin{gather}
D_{\mu}=1-r^{\mu}_{1}r^{\mu}_{2}e^{2iq_zl}.
\label{yysysys}
\end{gather}
%
%%%%%%%%%%%%%%%%%%%%%%%%%%%%%%%%%%%%%%%%%%%%%%%%%%%%%%%%%%%%%%%%%%%%%%%%%%%%%%%%%
\subsection{\label{sec:MMEssee}Green's function with the source in a body and the observation point in the vacuum gap}
%%%%%%%%%%%%%%%%%%%%%%%%%%%%%%%%%%%%%%%%%%%%%%%%%%%%%%%%%%%%%%%%%%%%%%%%%%%%%%%%%
The Fourier transform of the transmitted Green functions with  the
observation point ${\bf r}$ in the vacuum gap and the source point ${\bf r}'$ in the body $1$ or
$2$, are respectively  \cite{Tomas95}:
\begin{widetext}
\begin{eqnarray}
g_{ij}^{(1)}\left[\omega;{\bf Q},z,z'\right]&=&\frac{2\pi i k^2}{q_z^{(1)}}\sum_{\mu=s,p}\frac{t^{\mu}_{1}}{D_{\mu}}
\left[{ e}_{\mu,i}(+)\;{ e}_{\mu,j}^{(1)}(+)\;e^{iq_zz}+
{ e}_{\mu,i}(-)\;{ e}_{\mu,j}^{(1)}(+)\;r_{2}^{\mu}\;e^{-iq_zz}\;e^{2iq_zl}\right]\;e^{-iq_z^{(1)}z'},
\label{pppeeepeaap}\\
g_{ij}^{(2)}\left[\omega;{\bf Q},z,z'\right]&=&\frac{2\pi i k^2}{q_z^{(2)}}\sum_{\mu=s,p}\frac{t^{\mu}_{2}}{D_{\mu}}
\left[{ e}_{\mu,i}(-)\;{ e}_{\mu,j}^{(2)}(-)\;e^{-iq_zz}+
{ e}_{\mu,i}(+)\;{ e}_{\mu,j}^{(2)}(-)\;r_{1}^{\mu}\;e^{iq_zz}\right]\;e^{i q_zl}\;e^{iq_z^{(2)}(z'-l)}.
\label{pppeeepeaapss}
\end{eqnarray}
\end{widetext}
The symmetry of the problem becomes clear when one set the
origin of the coordinate axis in the center of the vacuum gap, by
changing $z\rightarrow z-l/2$ and $z'\rightarrow z'-l/2$ in
Eq.(\ref{pppeeepep}), (\ref{pppeeepeaap}) and (\ref{pppeeepeaapss}).
%
%%%%%%%%%%%%%%%%%%%%%%%%%%%%%%%%%%%%%%%%%%%%%%%%%%%%%%%%%%%%%%%%%%%%%%%%%%%%%%%%%
\section{\label{sec:GLevs}Force acting on a rarefied slab}
%%%%%%%%%%%%%%%%%%%%%%%%%%%%%%%%%%%%%%%%%%%%%%%%%%%%%%%%%%%%%%%%%%%%%%%%%%%%%%%%%
As discussed in Sec. \ref{AtSurfNeq}, in order to recover the
surface-atom force starting from the surface-surface expression, one
must consider the rarefied body as occupying a slab of finite
thickness. In this case, for non absorbing atom the PW term of the pressure is negligible, and the EW one
reproduces entirely the surface-atom force derived in  \cite{articolo2}. In this section we
discuss this problem and show explicitly that the PW term can be neglected. Let us consider the problem of the thermal
forces between a body $1$ at temperature $T$, which occupies the half-space ($z<0$), and a body $2$  at zero temperature which occupies a slab of thickness $L$ in the region ($l<z<l+L$). In the gap $0<z<l$ (region $0$)
and outside of the slab $z>l+L$ (region $3$) we can take $\varepsilon
=1$. The force per unit of area, acting on the slab in
$z-$direction, is
\begin{equation}
P\left( T,0\right) =P^{(0)}-P^{(3)}=\left\langle T_{zz}^{\left( 0\right)
}\right\rangle -\left\langle T_{zz}^{\left( 3\right) }\right\rangle,
\label{Pslab}
\end{equation}
where $T_{zz}^{\left( 0\right) }$ and $T_{zz}^{\left( 3\right) }$ are the $%
zz-$component of the Maxwell stress tensor in vacuum, calculated in the regions $0$ and $3$, respectively. For a
completely absorbing slab there is no field in the region $3$,
$T_{zz}^{\left( 3\right) }=0$ and one return to Eq.(\ref{force}). Of course from (\ref{Pslab}) one can calculate the
force acting on a slab of arbitrary thickness, and can recover the results of this paper relative to a thick slab. Here we we assume that
the slab is rarefied:
\begin{equation}
\varepsilon _{20}-1\ll 1\ .  \label{rar}
\end{equation}
Our goal will be to prove that for a slab without
absorption the propagating waves give the contribution $P^{{\rm
PW}}\ll P^{{\rm EW}}$, and hence can be neglected. For the proof it is
enough to consider a monochromatic component of the thermal
radiation impinging on the surface of the body 2 with the wave
vector ${\bf k}$ and polarization $\mu =s,p.$ In the terms of the
complex amplitudes of the fields its contribution to the pressure
can be written as [we omit $\left( \mu, {\bf k}\right) $ arguments
of the fields]
\begin{equation}
T_{zz}\left( \mu ,{\bf k}\right) =\frac{1}{8\pi }\left( \left| E_{z}\right|
^{2}-\frac{1}{2}\left| {\bf E}\right| ^{2}+\left| H_{z}\right| ^{2}-\frac{1}{%
2}\left| {\bf H}\right| ^{2}\right) .
\end{equation}
The fields in the region 0 are the sums of incident $\left(
+\right) $ and reflected $\left( -\right) $ waves:
\begin{equation}
{\bf E}^{\left( 0\right) }={\bf E}^{\left( 0+\right) }+{\bf E}^{\left(
0-\right) },\;\;\;\;\;{\bf H}^{\left( 0\right) }={\bf H}^{\left( 0+\right) }+{\bf H}%
^{\left( 0-\right) }{\rm \ ,}
\end{equation}
where ${\bf E}^{\left( 0+\right) },{\bf H}^{\left( 0+\right) }\propto
e^{iq_{z}z}$ and ${\bf E}^{\left( 0-\right) },{\bf H}^{\left( 0-\right)
}\propto e^{-iq_{z}z}$. An important point of the proof is that incident and
reflected waves give independent contributions to the stress tensor:
\begin{equation}
T_{zz}^{\left( 0\right) }=T_{zz}^{\left( 0+\right) }+T_{zz}^{\left(
0-\right) }.  \label{add}
\end{equation}
The additivity property (\ref{add}) is obvious. Presence of the
mixed term containing both ${\bf E}^{\left(
0+\right) }$ and ${\bf E}^{\left( 0-\right) \ast }$
would result in the $z-$dependence of $T_{zz}$. But this is not possible
since it violates the momentum conservation.\\
By definition we have that
\begin{equation}
\left| {\bf E}^{\left( 0-\right) }\left( \mu, {\bf k}\right) \right|
^{2}=R^{\left( \mu ,{\bf k}\right) }\left| {\bf E}^{\left( 0+\right) }\left(
\mu ,{\bf k}\right) \right| ^{2},
\end{equation}
where $R^{\left( \mu, {\bf k}\right) }$ is the reflection coefficient from the slab, for the $%
(\mu, {\bf k})$-wave . Taking into account the Fresnel relations
between the field components at the reflection, we easily find that
\begin{equation}
T_{zz}^{\left( 0-\right) }=RT_{zz}^{\left( 0+\right) },\;\;\;\;\;\;\;\;
T_{zz}^{\left( 0\right) }=\left( 1+R\right) T_{zz}^{\left( 0+\right)
}.
\end{equation}
Let us consider now the fields in the vacuum region 3. There is only a refracted
wave and we have $\left| {\bf E}^{\left( 3\right) }\right| ^{2}=D^{\left(
\mu, {\bf k}\right) }\left| {\bf E}^{\left( 0+\right) }\right| ^{2}$, where \
$D^{\left( \mu, {\bf k}\right) }$ is the transmission coefficient. In 
absence of absorption $D^{\left( \mu, {\bf k}\right) }=1-R^{\left( \mu, {\bf k}%
\right) }$. This means that
\begin{eqnarray}
T_{zz}^{\left( 3\right) }\left( \mu, {\bf k}\right) &=&\left( 1-R^{\left( \mu,
{\bf k}\right) }\right) T_{zz}^{\left( 0+\right) }\left( \mu, {\bf k}\right) \notag\\
&=&\frac{1-R^{\left( \mu, {\bf k}\right) }}{1+R^{\left( \mu, {\bf k}\right) }}%
T_{zz}^{\left( 0\right) }\left( \mu, {\bf k}\right),
\end{eqnarray}
and from (\ref{Pslab}) one has
\begin{equation}
P^{{\rm PW}}\left( \mu, {\bf k}\right) =\frac{2R^{\left( \mu, {\bf k}\right) }%
}{1+R^{\left( \mu, {\bf k}\right) }}T_{zz}^{\left( 0\right) }\ \left( \mu,
{\bf k}\right) .
\label{Final}
\end{equation}
One can easily calculate $R^{\left( \mu, {\bf k}\right) }$, (see,
for example, the problem N.4 in \S\ 66\ of \cite{LLPCM}). At real
$\varepsilon _{20}\rightarrow 1$ one gets, independently on the
polarization, the result:
\begin{equation}
R^{\left( \mu, {\bf k}\right) }\approx \frac{\sin ^{2}\left[ \frac{\omega L}{c%
}\cos \theta _{0}\right] }{4\cos ^{4}\theta _{0}}\left( \varepsilon
_{20}-1\right) ^{2}  ,
\label{R}
\end{equation}
where $\theta _{0}$ is the angle of
incidence. This equation is valid at
the condition $\cos \theta _{0}\gg \sqrt{ \varepsilon _{20}-1}\ .$
Let us note that the surface-atom force equations of
\cite{articolo2} must be valid in the ''additive`` regime of the Sec.
\ref{sec:LongDnonad}, where 
just the incident angles $\cos \theta _{0}=q_{z}/k\sim \lambda
_{T}/l$ $\gg \sqrt{\varepsilon _{20}-1}$ \ are important [see Eq.(\ref{additivity})]. For such
angles $R^{\left( \mu, {\bf k}\right) }\sim \left(l \sqrt{
\varepsilon _{20}-1}/\lambda _{T}\right) ^{4}\ll 1$, and $P^{{\rm
PW}}\ll T_{zz}^{\left( 0\right) }$. Here we assumed that $l$
$\gg \lambda _{T} $. \ For $l$ $\lesssim \lambda _{T}$ one gets
simply $R^{\left( \mu, {\bf k} \right) }\sim \left(
\varepsilon _{20}-1\right) ^{2}$. It is not difficult to check that
$T_{zz}^{\left( 0\right) }\sim $ $\left( \varepsilon _{20}-1\right)
\sim P^{{\rm EW}}.$ Finally we find that $P^{{\rm PW}}\ll P^{ {\rm
EW}}$ and hence the propagating waves contribution can be neglected.

Let us discuss now the role of a weak absorption. Consider the case
\begin{equation}
\varepsilon _{2}=\varepsilon _{2}^{\prime }+i\varepsilon _{2}^{\prime \prime
},\varepsilon _{2}^{\prime \prime }\ll 1,\varepsilon _{2}^{\prime }\approx 1.
\end{equation}
It is not difficult to generalize (\ref{Final}) for a slab with absorption:
\begin{equation}
P^{{\rm PW}}=\left( 1-\frac{D}{1+R}\right) T_{zz}^{\left( 0\right)
}, \label{abs}
\end{equation}
where the transmission coefficient $D<1-R$. If $R\ll 1,$
\begin{equation}
P^{{\rm PW}}\approx \left( 1-D\right) T_{zz}^{\left( 0\right) }.
\end{equation}
According to Problem 4 of \S\ 66\ in \cite{LLPCM}, one has that
$D\sim \exp \left( -\frac{\omega L\varepsilon _{2}^{\prime \prime
}}{c\cos \theta _{0}}\right) .$ The imaginary part \ $\varepsilon
_{2}^{\prime \prime }\left( \omega \right) $ must be taken in this
estimate at $\omega \sim k_BT/\hbar .$ The factor $\left( 1-D\right)
$ and correspondingly $P^{{\rm PW}}$ are small if $\omega
L\varepsilon _{2}^{\prime \prime }\ll c\cos \theta _{0}$. This gives
the condition for neglecting the absorption:
\begin{equation}
\varepsilon _{2}^{\prime \prime }\left( \omega \sim k_B T/\hbar \right) \ll
\lambda _{T}^{2}/lL.  \label{condD}
\end{equation}
Let us note that for evanescent waves $T_{zz}^{\left( 3\right){\rm EW} }$ does not depend on $z$, while the field of an
evanescent wave goes to zero at $z\rightarrow \infty $. This means that $T_{zz}^{\left( 3\right){\rm EW} }\equiv 0$.
%
%%%%%%%%%%%%%%%%%%%%%%%%%%%%%%%%%%%%%%%%%%%%%%%%%%%%%%%%%%%%%%%%%%%%%%%%%%%%%%%%%%%%%%%%%%%%%%%%

\end{document}